%% file: JossyGupta_2023.tex
\documentclass[%
 aip,
 amsmath,amssymb,
preprint,%
]{revtex4-1}

\usepackage{graphicx}
\usepackage{dcolumn}
\usepackage{bm}

\usepackage[utf8]{inputenc}
\usepackage[T1]{fontenc}
\usepackage{mathptmx}
\usepackage{etoolbox}
\usepackage{notoccite}
\usepackage{xcolor}
\usepackage{multirow}
\usepackage{changes}
\DeclareMathAlphabet{\mathcal}{OMS}{cmsy}{m}{n}
\makeatletter
\def\@email#1#2{%
 \endgroup
 \patchcmd{\titleblock@produce}
  {\frontmatter@RRAPformat}
  {\frontmatter@RRAPformat{\produce@RRAP{*#1\href{mailto:#2}{#2}}}\frontmatter@RRAPformat}
  {}{}
}%
\makeatother
\input{./defs}

\newif\ifmarkedup
\markedupfalse
\ifmarkedup
\newcommand{\Revision}[2]{\replaced{#2}{#1}}
\else
\newcommand{\Revision}[2]{{\color{black}#2}}
\fi

\newif\ifmarkedupTwo
\markedupTwofalse
\ifmarkedupTwo
\newcommand{\RevisionTwo}[2]{\replaced{#2}{#1}}
\else
\newcommand{\RevisionTwo}[2]{{\color{black}#2}}
\fi
\begin{document}

\preprint{Preprint submitted to Physics of fluids on February 28, 2023}

\title{Baroclinic interaction of forced shock waves with random thermal gradients }
\author{Joaquim P. Jossy}
\author{Prateek Gupta}
\email{prgupta@iitd.ac.in}
 \homepage{https://web.iitd.ac.in/~prgupta/}
 
\affiliation{ 
Department of Applied Mechanics, Indian Institute of Technology Delhi, New Delhi 110016, India 
}%

\date{\today}

\begin{abstract}
Density gradients aligned at an angle to pressure gradients result in baroclinic torque in fluid flows, generating vorticity. In this work, we study the vorticity generated by the baroclinic torque exerted by the interaction of pressure jumps across random two-dimensional shock waves with density gradients. A field of random two-dimensional shock waves has acoustic spectral energy scaling as $\widehat{E}_k \sim \varepsilon^{2/3}\ell^{-1/3}k^{-2}$ where $k$ is the wavenumber, $\varepsilon$ is the energy dissipation, and $\ell$ is the integral length scale of the field. Since the acoustic energy is broadband, pressure and velocity gradients exist in a wide range of length scales. We study the interaction of these broadband gradients with isobaric thermal gradients localized at a length scale in the spectral space. We show that the method of generating shock waves or injection of wave energy in the system governs the baroclinic interactions. For stochastically forced shock waves, baroclinic terms are negligible. Broadband vorticity with energy at least two orders of magnitude smaller is generated due to continuous variation in curvature of shock waves caused by stochastic forcing. On the other hand, shock waves maintained by energy rescaling result in the generation of coherent vorticity. We also discuss the relative magnitude of the baroclinic torque generated due to total density gradients compared to the \Revision{baroclinic torque}{one} generated due to non-isentropic density gradients within the shock waves interacting with the pressure gradients. 
\end{abstract}
\newcommand{\PG}[1]{{\textbf{\color{red}{#1}}}}
\newcommand{\Jo}[1]{{\textbf{\color{blue}{#1}}}}

\maketitle

\section{Introduction}
Finite amplitude or nonlinear acoustic waves exist in various engineering and scientific applications such as aerospace~\cite{goldstein1990spatial, liepmann2001elements,baars2013transient, bonciolini2017output, gupta2017spectral}, nondestructive testing~\cite{breazeale1963finite, lissenden2021nonlinear}, astrophysics \cite{sagdeev19791976}, nuclear energy~\cite{brouillette2002richtmyer}, and medicine~\cite{lukes2016tandem}. Nonlinear effects such as acoustic streaming~\cite{lighthill1978acoustic, shimizu2010numerical} and wave steepening~\cite{whitham2011linear,gupta2018spectral, thirani2020knudsen} govern the propagation of such finite amplitude nonlinear acoustic waves in a compressible fluid. Due to steepening, nonlinear acoustic waves form propagating shock waves~\cite{whitham2011linear, gupta2018spectral}. Propagation, dissipation, and dispersion of these shock waves can be altered by the thermodynamic properties of the medium in which they propagate. Furthermore, interaction of shock waves with inhomogeneities in the medium may generate additional hydrodynamic quantities of interest such as vorticity. In this work, we study the propagation of forced shock waves in an inhomogeneously heated medium in two dimensions. 

~\citet{ellermeier1993nonlinear} showed that non-uniform cross-section and density stratification affect the nonlinear distortion of planar waves.~\citet{tyagi2003nonlinear} found that entropy gradients adversely affect the nonlinear wave steepening and shock formation.~\citet{prasad2006weakly} studied one-dimensional shock waves in slowly varying isentropic one-dimensional flows. Using multiple scales analysis,~\citet{prasad2006weakly} examined the effect of non-uniform flow in a duct on evolution and transport of acoustic power in the duct and found that the mean flow has an impact on the residence time of shock waves in the duct due to viscous dissipation inside the shock waves. In this work, we focus on the propagation of shock waves in a quiescent thermally inhomogeneous medium in two dimensions. Due to the misalignment of pressure gradients across the shock waves and density gradients, baroclinic torque generates vorticity. Such misalignment may exist due to the thermoviscous dissipation inside the shock waves or externally imposed thermal gradients. The vorticity caused by such baroclinic interaction can be used to enhance fuel-air mixing in air-breathing engines~\cite{budzinski1992rayleigh, andreopoulos2000shock, romagnosi2011role, tian2017numerical, wong2022analysis} or active flow-control~\cite{singh2019experimental}. We perform shock-resolved direct numerical simulations (DNS) of two-dimensional randomly generated shock waves and study their interaction with background thermal gradients. We focus on the length scales and magnitude of vorticity generated due to baroclinic interaction between pressure gradients across the shock waves and background density stratification generated by thermal gradients. 

In baroclinic flows, fluid density is a function of pressure, temperature, and composition or concentration of the dissolved constituents \cite{kundu2015fluid}. This dependency can create a misalignment in the direction of the density and pressure gradients\Revision{ }{, unlike barotropic flows in which the density and pressure gradients are parallel}. The resulting torque due to the misalignment of density and pressure gradients is called baroclinic torque and \Revision{is}{may} \Revision{responsible for increasing or decreasing}{result in increase or decrease of} the vorticity and circulation in the fluid\Revision{}{, depending on the existing vorticity field}. In inertially confined fusion reactors, baroclinic torque results in Richtmyer-Meshkov instability~\cite{mikaelian1985richtmyer, brouillette2002richtmyer} which causes mixing of the fuel and capsule material limiting the efficiency of the reactor. In supersonic combustion chambers, shock-flame interaction results in vorticity generated due to the resulting baroclinic torque which is argued as the primary cause of deflagration to detonation transition~\cite{yang2021dynamics}. \Revision{}{Additionally, baroclinic production is attributed as the primary mechanism of enhanced mixing downstream of an oblique shock wave as streamwise vorticity passes through it~\cite{yu2020two, wei2022flow}. }Usually, only stratification or density gradient is considered while analyzing the baroclinic torque. However, in a multi-dimensional field of shock waves propagating in a homogeneous medium, the density gradient is not necessarily parallel to the pressure gradient due to local entropy generation. Consequently, the interaction of pressure gradients with density gradients in a two-dimensional or three-dimensional field of shock waves is possible without prescribed or background stratification as well. In this work, we analyze the interaction of two-dimensional field of random shock waves with random background thermal gradients using two-dimensional shock-resolved direct numerical simulation (DNS) of fully compressible Navier-Stokes equations. Such background thermal gradients result in the stratification of the fluid in which the shock waves propagate. We generate these shock waves using two contrasting forcing methods. In one method, we force the momentum equations using a stochastic process, also used to generate equilibrium turbulence~\cite{eswaran1988examination, jagannathan2016reynolds}. In the other method, we rescale the density-weighted wave spectral energy in the system in a band of small wavenumbers (large length scales). In the latter forcing method, energy loss to smaller length scales due to acoustic spectral energy cascade~\cite{gupta2017spectral, gupta2018spectral} is compensated at each time step of the simulation. Furthermore, the relative phasing of shock waves is maintained since only amplitudes are rescaled. Using numerical simulations, we show that the vorticity generated due to different forcing methods is at different length scales. For random forcing, the vorticity and enstrophy are broadband, while they are coherent for energy rescaling. For energy rescaling, we discuss the enstrophy budgets to elucidate the mechanism of coherent vorticity generation in the system. \Revision{}{To the best of our knowledge, this is the first study in which the interaction of random shock waves with a inhomogeneously stratified medium has been studied.}

In Section~\ref{sec: governing_equations} we discuss the governing equations of our simulations and the \Revision{ }{theoretical} derivation of \Revision{}{wave} spectral energy equation using second-order \Revision{}{nonlinear} perturbation equations. \Revision{}{Since vorticity is expected to be generated due to baroclinic interactions, we also discuss the decomposition of flow fields in rotational and dilatational components to isolate the effects on the spectral energy dynamics of waves.} In Section~\ref{sec: Numerical_simulations} we discuss the numerical setup and problem formulation. Particularly, we outline the two different forcing methods used to maintain equilibrium with shock waves without any preference of direction of propagation. We also show that for simulations with no background thermal gradients, density wave spectral energy $\widehat{E}^f_{wk}$  scales as $\varepsilon^{2/3}\ell^{-1/3} k^{-2}$ (see Section~\ref{sec: Numerical_simulations} for definitions). In Section~\ref{sec: resultsDiscussion} we discuss the length scales of the two-dimensional field of random shock waves generated due to forcing, the magnitude and length scales of vorticity generated, and the enstrophy budgets before concluding in Section~\ref{sec: conclusions}.

\section{Flow field decomposition and spectral energy transfer}
\label{sec: governing_equations}
In this section, we discuss the \Revision{ }{theoretical background and the} derivation of nonlinear governing equations for perturbations in density, velocity\Revision{field}{}, and pressure \Revision{}{field} from fully compressible Navier-Stokes equations for an ideal gas. Using these nonlinear perturbation equations, we derive the expressions for the flux of spectral wave energy and spectral dissipation. To this end, we discuss the possible decomposition of fields into the wave and the vortical components.

Dimensionless fully compressible Navier-Stokes equations for an ideal gas are given by, 
\begin{subequations}
 \begin{align}
       \frac{\partial \rho}{ \partial t} + \nabla \cdot (\rho \bfu) =0 ,
       \label{eq: continuity}
 \end{align}
 \begin{align}
 \frac{\partial \bfu}{\partial t} + \bfu \cdot \nabla \bfu + \frac{\nabla p}{\rho} = \frac{1}{ \rho Re_{ac} }  \left( \nabla \cdot( 2 \mu \bfS )\right) +\nonumber \\
  \frac{1}{ \rho Re_{ac} }  \left( \nabla \cdot \left( \kappa - \frac{2 \mu }{3} \right) \bfD \right) + \bfF ,
  \label{eq: momentum_equation} 
  \end{align}
 \begin{align}
\frac{\partial p}{\partial t} + \bfu \cdot \nabla p + \gamma p\nabla \cdot \bfu =  \frac{1}{Re_{ac} \hspace{0.8mm} \mathrm{Pr}} \nabla \cdot (\alpha \nabla T) +\nonumber \\ \frac{\gamma -1 }{Re_{ac}}\left(2 \mu \bfS:\bfS +\left( \kappa-\frac{2 \mu}{3}\right)\bfD:\bfD   \right) ,
 \label{eq: pressure_equation}
 \end{align}
\end{subequations}
combined with the dimensionless ideal gas equation of state, 
\begin{equation}
 \gamma p = \rho T.\label{eq: ideal_gas_nd}
\end{equation}
In Eqs.~\eqref{eq: momentum_equation} and~\eqref{eq: pressure_equation}, $\bfS$ and $\bfD$ are the strain rate and dilatation tensors, respectively, given by, 
\begin{equation}
 \bfS = \frac{1}{2}\left(\nabla\bfu + \nabla\bfu^T\right), ~\bfD = \nabla\cdot\bfu \bfI.
\end{equation}
In Eq.~\eqref{eq: momentum_equation}, $\bfF$, is the external force vector which is used to maintain random shocks in a two-dimensional domain and $\kappa$, $\mu$, $\alpha$, $ Re_{ac} $, and Pr are the dimensionless bulk viscosity, dimensionless dynamic viscosity, dimensionless thermal conductivity, acoustic Reynolds number, and Prandtl number, respectively. As discussed in detail in next section, the external force vector $\bfF$ may represent the stochastic forcing or the energy rescaling. Furthermore, to obtain the dimensionless Eqs.~\eqref{eq: continuity},~\eqref{eq: momentum_equation}, and \eqref{eq: pressure_equation}, following scales have been used, 
\begin{equation}
 [\rho] = \rho_m,~[p] = \gamma p_m,~[T] = T_m,~[\bfu] = (c_m, c_m),~[x,y] = (L, L).
 \label{eq: referenceScales}
\end{equation}
Subscript $()_m$ in above relations denotes characteristic values corresponding to the quiescent medium. Velocity scale $c_m$ denotes the speed of sound at these characteristic values and $L$ corresponds to a length scale. Using scales in Eq.~\eqref{eq: referenceScales}, $Re_{ac}$ is defined as, 
\begin{equation}
 Re_{ac} = \frac{\rho_m c_m L}{\mu_m},
\end{equation}
where $\mu_m$ is dynamic viscosity at $T_m$. Throughout this work, we analyse results from numerical simulations of dimensionless equations. Hence, values of scales in Eqs.~\eqref{eq: referenceScales} are irrelevant and only values of dimensionless numbers are specified.

In this work, we study the interaction of random nonlinear acoustic fields (weak shock waves~\cite{gupta2018spectral}\Revision{ }{, also see results in Section~\ref{sec: simulation_parameters}}) with random thermal gradients. To this end, we perform DNS of Eqs.~\eqref{eq: continuity}-\eqref{eq: pressure_equation} (see Section~\ref{sec: Numerical_simulations}) and extract relevant physical quantities using the second order nonlinear governing equations for perturbations in density, velocity, and pressure. To derive these equations, we consider a quiescent base state with isobaric thermal gradients. The fields of dependent variables $\bfu, p, \rho,$ and $T$ can be written as a decomposition of the base field and the random perturbations denoting the random nonlinear acoustic fields,
\begin{equation}
 \bfu = \bfu',~p = \frac{1}{\gamma} + p',~\rho = f(x,y) + \rho',
\end{equation}
and 
\begin{equation}
T = \frac{1}{f(x,y)} + T'.
\label{eq: thermal_gradient_init}
\end{equation}
Field $1/f(x,y)$ corresponds to the base state thermal gradients (which may be interpreted as base state stratification as well). \Revision{}{For a homogeneous medium, $f(x,y)=1$}. As we discuss in Section~\ref{sec: Numerical_simulations}, we initialize these thermal gradients within bands of wavenumbers which generates an inhomogeneous medium in which the random shock waves are forced and maintained at statistical steady state. \Revision{We also note that for homogeneous medium $f(x,y) = 1$}{}.

To understand energy cascade and interaction with background thermal gradients, we first derive nonlinear governing equations in $p'$ and $\bfu'$. Since the thermoviscous terms can only have second order contribution at the leading order (see Gupta and Scalo~\cite{gupta2018spectral} for a detailed discussion), we obtain the following dimensionless nonlinear equations correct up to second order,
\begin{subequations}
    \begin{align}
  \frac{\partial \bfu'}{\partial t} + \bfu' \cdot \nabla \bfu' + \frac{\nabla p'}{f} - \frac{p'\nabla p'}{f^{2}} = \nonumber \\
  \frac{1}{ f Re_{ac} }  \left ( \nabla \cdot \left (2 \mu \bfS' \right) +
\nabla \cdot\left ( \left ( \kappa - \frac{2 \mu}{3} \right ) \bfD'\right )  \right ) ,
\label{eq: momentum_nonlinear}
\end{align}
\begin{align}
    \frac{\partial p'}{\partial t} + \bfu' \cdot \nabla p' + \nabla \cdot u'+ \gamma p'\nabla \cdot \bfu' = \frac{1}{Re_{ac} \hspace{0.8mm} \mathrm{Pr}} \nabla \cdot (\alpha \nabla T') .\label{eq: pressure_nonlinear}
\end{align}
\end{subequations}
In Eq.~\eqref{eq: pressure_nonlinear}, temperature perturbation $T'$ is related to pressure and density perturbations ($p'$ and $\rho'$, respectively) via the dimensionless ideal gas equation of state (see Eq.~\eqref{eq: ideal_gas_nd}), 
\begin{equation}
 \gamma p' = \left(f + \rho'\right)\left(\frac{1}{f} + T'\right) = 1 + \frac{\rho'}{f} + fT' + \rho' T'.
\end{equation}
Density perturbation $\rho'$ is governed by, 
\begin{equation}
 \frac{\partial \rho'}{\partial t} + \nabla\cdot\left(f\bfu'\right) + \nabla\cdot\left(\rho'\bfu'\right) = 0.
\end{equation}

A spectral energy equation is possible for perturbation energy only if a true energy corollary can be derived for the governing system of equations~\cite{gupta2018spectral}. To formulate the spectral energy equation from Eqs.~\eqref{eq: momentum_nonlinear} and~\eqref{eq: pressure_nonlinear}, we \RevisionTwo{first note that}{highlight that}
the linearized governing equations for inviscid acoustic waves in a stratified medium, 
\begin{align}
 \frac{\partial \bfu'}{\partial t} + \frac{\nabla p'}{f} = 0\label{eq: linear_inviscid_u}~\mathrm{and}\\
 \frac{\partial p'}{\partial t} + \nabla\cdot\bfu'=0,
 \label{eq: linear_inviscid_p}
\end{align}
do not admit a modal decomposition for non-uniform $f$. For a homogeneous background medium ($f(x,y) = 1$ everywhere), the modal decomposition of the linearized Eqs.~\eqref{eq: linear_inviscid_u} and~\eqref{eq: linear_inviscid_p} is given by, 
\begin{equation}
\left(u', v', p'\right)^T = \sum_n\widehat{\bfphi}_{\pm} e^{(i(k_{nx} x + k_{ny} y \pm \omega_n t)} + \sum_n\widehat{\bfphi}_{0} e^{(i(k_{nx} x + k_{ny} y)}.
\end{equation}
where $u'$ and $v'$ are components of velocity field perturbation $\bfu'$. Modes $\widehat{\bfphi}_0$ and $\widehat{\bfphi}_{\pm}$ are given by,
\begin{align}
 \widehat{\bfphi}_0 = \frac{1}{|\bfk|}\left(-k_{ny}, k_{nx}, 0\right)^T,\\
 \widehat{\bfphi}_{\pm} = \frac{1}{|\bfk|\sqrt{2}}\left(k_{nx}, k_{ny}, \pm |\bfk| \right)^T,
\end{align}
where $\widehat{\bfphi}_0$ represents the vortical mode which has no contributions from the pressure perturbations and $\widehat{\bfphi}_{\pm}$ represent the acoustic modes. Since $\widehat{\bfphi}_0, \widehat{\bfphi}_{\pm}$ are orthogonal to each other, any two-dimensional compressible flow field can be projected on $\widehat{\bfphi}_0, \widehat{\bfphi}_{\pm}$. For second order nonlinear Eqs.~\eqref{eq: momentum_nonlinear}-~\eqref{eq: pressure_nonlinear}, writing the instantaneous solution as an expansion of $\widehat{\bfphi}_0, \widehat{\bfphi}_{\pm}$, we obtain, 
\begin{align}
 \left(u', v', p'\right)^T = &\sum_n\Big(\widehat{\Omega}(\bfk_n, t)\widehat{\bfphi}_0 \nonumber \\
 &+ \widehat{R}_+(\bfk_n,t)\widehat{\bfphi}_+ + \widehat{R}_-(\bfk_n,t)\widehat{\bfphi}_-\Big)e^{i(k_{nx} x + k_{ny} y)}.
 \label{eq: projection}
\end{align}
Here $\widehat{\Omega}(\bfk_n,t) $ corresponds to the vortical contribution of the perturbation field and $\widehat{R}_+(\bfk_{n}, t), $ and $\widehat{R}_-(\bfk_{n},t)$ correspond to the acoustic contribution. Equation~\eqref{eq: projection} yields a spectral energy density definition (since the modes $\widehat{\bfphi}_0, \widehat{\bfphi}_{\pm}$ are orthogonal) as,
\begin{equation}
 \widehat{E}_{\bfk}(t) = \frac{1}{2}|\widehat{\Omega}(\bfk_n,t)|^2 + \frac{1}{2}|\widehat{R}_+(\bfk_n,t)|^2 + \frac{1}{2}|\widehat{R}_-(\bfk_n,t)|^2.
\end{equation}
Since there is no such modal decomposition possible for waves in an inhomogeneous medium, one may follow~\citet{miura1995acoustic} and use the spatial Fourier modes of density-weighted velocity perturbations $\bfw' = \sqrt{f}\bfu'$ to define the spectral energy as,
\begin{align}
 \widehat{E}_{\bfk} &= \frac{1}{2}\left(\widehat{\bfw}^*_{\bfk}\cdot\widehat{\bfw}_{\bfk} + \widehat{p}^*_{\bfk}\widehat{p}_{\bfk}\right)\nonumber \\
 &= \underbrace{\frac{1}{2}\left(\widehat{\bfw}^*_{v\bfk}\cdot\widehat{\bfw}_{v\bfk}\right)}_{\widehat{E}^f_{v\bfk}} + \underbrace{\frac{1}{2}\left( \widehat{\bfw}^*_{w\bfk}\cdot\widehat{\bfw}_{w\bfk}+ \widehat{p}^*_{\bfk}\widehat{p}_{\bfk}\right)}_{\widehat{E}^f_{w\bfk}},
\end{align}
where $()^*$ denotes complex conjugate, $\widehat{\bfw}_{\bfk}$ is the Fourier transform of $\bfw'$ and $\bfw_{v{\bfk}}, \bfw_{w{\bfk}}$ are obtained using the Helhmholtz decomposition in the Fourier space, 
\begin{subequations}
\begin{equation}
   \widehat{\bfw}_{\bfk} = \widehat{\bfw}_{v{\bfk}} + \widehat{\bfw}_{w{\bfk}},\\
\end{equation}
\begin{equation}
  i\bfk\cdot  \widehat{\bfw}_{\bfk} = i\bfk\cdot\widehat{\bfw}_{w{\bfk}},~\mathrm{and}\\
\end{equation}
\begin{equation}
  i\bfk\times\widehat{\bfw}_{\bfk} = i\bfk\times\widehat{\bfw}_{v{\bfk}}.
\end{equation}
 \label{eq: Decomposition}
\end{subequations}
Energy can be decomposed as $\widehat{E}^f_{w\bfk}$ and $\widehat{E}^f_{v\bfk}$ where the superscript $()^f$ denotes energy evaluated using Fourier modes of $\bfw'$. However, as we show in further sections, $E^f_{v\bfk}$ is not a good choice for quantifying the actual vortical energy in the system corresponding to the local rotation rate of the fluid. Hence, we reserve $\widehat{E}_{v\bfk}$ for the \emph{actual vortical energy} defined as, 
\begin{equation}
    \widehat{E}_{v\bfk} = \frac{1}{2}\widehat{\bfu}^*_{v\bfk}\cdot\widehat{\bfu}_{v\bfk},
\end{equation}
where $\widehat{\bfu}_{v\bfk}$ satisfies, 
\begin{equation}
    i\bfk\times\widehat{\bfu}_{v\bfk} = i\bfk\times\widehat{\bfu}_{\bfk}.
    \label{eq: rotational_velocity}
\end{equation}
In sections below, we use $\widehat{E}^f_{wk}$ to represent the spectral energy in waves. In the physical space, perturbation energy equation can be written as, 
\begin{align}
 \frac{\partial }{\partial t} \left(\frac{f|\bfu'|^2}{2} + \frac{p'^2}{2}\right) + \nabla\cdot\left(\bfu'p'\right) + f\bfu'\cdot\left(\bfu'\cdot\nabla\bfu'\right) + \nonumber\\
 \gamma p'^2\nabla\cdot\bfu'  = f\bfu'\cdot D_u + p'D_p,
 \label{eq: perturbation_space}
\end{align}
where $D_u$ and $D_p$ correspond to the thermoviscous dissipation terms from Eq.~\eqref{eq: momentum_nonlinear} and Eq.~\eqref{eq: pressure_nonlinear}, respectively. Equation~\eqref{eq: perturbation_space} is obtained by taking the dot product of Eq.~\eqref{eq: momentum_nonlinear} with $f\bfu'$ and Eq.~\eqref{eq: pressure_nonlinear} with $p'$.

As shown by~\citet{gupta2018spectral}, the $\gamma p'^2\nabla\cdot\bfu'$ term can be further decomposed into an infinite series expansion to yield an exact energy corollary for planar one-dimensional nonlinear acoustic waves. However, such a decomposition is hard in higher dimensions even for waves propagating in a homogeneous medium. Consequently we choose to work with the remaining terms for obtaining the spectral energy cascade terms. In spectral space, the energy equation can be represented as, 
\begin{equation}
 \frac{d \widehat{E}_{\bfk}}{dt} + \widehat{\mathcal{T}}_{\bfk} = \widehat{\mathcal{D}}_{\bfk},
\end{equation}
where $\widehat{\mathcal{T}}_{\bfk}$ is the spectral energy transfer function and $\widehat{\mathcal{D}}_{\bfk}$ is the spectral dissipation function. From numerical simulations, we use discretized spectral quantities ($\widehat{E}_{\bfk}, \widehat{\mathcal{T}}_{\bfk}, \widehat{\mathcal{D}}_{\bfk}$ and derived quantities) to investigate spectral energy dynamics using the corresponding binned quantities as, 
\begin{equation}
 \widehat{\Phi}_k = \sum^{|\bfk| < k + \Delta k/2}_{|\bfk| > k - \Delta k/2}\Phi_{\bfk} = b(\widehat{\Phi}_{\bfk}),
\end{equation}
where $\Phi$ may denote spectral energy, spectral energy transfer, spectral dissipation or derived quantities. Binned spectral energy transfer and spectral dissipation are obtained by multiplying Eq.~\eqref{eq: momentum_nonlinear} with $\sqrt{f}$ and then taking the dot-product with $\widehat{w}^*_{wk}$ in the spectral space (and similarly multiplying Eq.~\eqref{eq: pressure_nonlinear} with $\widehat{p}^*_k$ in the spectral space) as,
\begin{align}
 \widehat{\mathcal{T}}_{k} = b\Bigg(\mathrm{Re}\left[\widehat{\bfw}^*_{\bfk}\cdot\left(\widehat{\frac{\nabla p'}{\sqrt{f}}}\right)_{\bfk} + \widehat{p}^*_{\bfk}\widehat{\nabla\cdot\bfu'}_{\bfk}\right] + \nonumber \\
 \mathrm{Re}\left[\widehat{\bfw}^*_{\bfk}\cdot\left(\widehat{\bfw\cdot\nabla\bfu'}\right)_{\bfk} - \widehat{\bfw}^*_{\bfk}\cdot\left(\widehat{\frac{p'\nabla p'}{\sqrt{f}}}\right)_{\bfk}\right] + \nonumber \\
 \mathrm{Re}\left[\gamma\widehat{p}^*_{\bfk}\left(\widehat{p'\nabla\cdot\bfu'}\right)_{\bfk} + \widehat{p}^*_{\bfk}\left(\widehat{\bfu'\cdot\nabla p'}\right)_{\bfk}\right]\Bigg),
 \label{eq: transfer_function_f}
\end{align}
and
\begin{equation} \label{eq1}
\begin{split}
 \widehat{\mathcal{D}}_{k} = b\Bigg(\mathrm{Re}\left[ \widehat{\bfw}^*_{\bfk}\cdot \left(\widehat{\frac{\mu}{\sqrt{f}\hspace{0.8mm} Re_{ac}} \left(\nabla\cdot (\nabla u^{'} + \nabla u'^{T} )\right)}\right)_{\bfk}
 \right] + \\
 \mathrm{Re}\left[ \widehat{\bfw}^*_{\bfk}\cdot \left(\widehat{\frac{1}{\sqrt{f}\hspace{0.8mm} Re_{ac}} \nabla\cdot\left( \kappa -\frac{2 \mu}{3}(\nabla \cdot u') )\right)}\right)_{\bfk}
 \right] + \\ 
\mathrm{Re}\left[ \widehat{\bfp}^*_{\bfk}\cdot \left(\widehat{\frac{1}{Re_{ac}\hspace{0.8mm}\mathrm{Pr} } \nabla \cdot (\alpha \nabla T')} \right)_{\bfk}
 \right]\Bigg),
\end{split}
\end{equation}
respectively. For uniform background temperature and density ($f=1$), Eq.~\eqref{eq: transfer_function_f} yields the transfer function for a nonlinear acoustic system in a homogeneous medium, 
\begin{align}
  \widehat{\mathcal{T}}_{k}= b\Bigg(\mathrm{Re}\left[\widehat{\bfu}^*_{\bfk}\cdot\left(\widehat{\bfu\cdot\nabla\bfu'}\right)_{\bfk} - \widehat{\bfu}^*_{\bfk}\cdot\left(\widehat{p'\nabla p'}\right)_{\bfk}\right] + \nonumber \\
 \mathrm{Re}\left[\gamma\widehat{p}^*_{\bfk}\left(\widehat{p'\nabla\cdot\bfu'}\right)_{\bfk} + \widehat{p}^*_{\bfk}\left(\widehat{\bfu'\cdot\nabla p'}\right)_{\bfk}\right]\Bigg).
\end{align}

Throughout this work, we quantify the results from numerical simulations of fully nonlinear governing Eqs.~\eqref{eq: continuity}-\eqref{eq: pressure_equation} (discussed in Section~\ref{sec: Numerical_simulations}) using second order nonlinear Eqs.~\eqref{eq: momentum_nonlinear} and~\eqref{eq: pressure_nonlinear} and corresponding spectral energy cascade quantities derived.
\section{Numerical Simulations}
\label{sec: Numerical_simulations}
In this section we discuss the computational setup for our numerical simulations of two-dimensional fully compressible dimensionless Navier-Stokes Eqs.~\eqref{eq: continuity}-\eqref{eq: pressure_equation}. 
We solve Eqs.~\eqref{eq: continuity}-\eqref{eq: pressure_equation} numerically using the fourth order Runge-Kutta time stepping and standard Fourier pseudospectral method~\cite{boyd2001chebyshev} combined with $2/3$ deliasing for numerical stability and slab-decomposition for parallel distributed computing using MPI~\cite{mortensen2016high}. \Revision{All the simulations presented in this work correspond to a grid of $3072\times 3072$ in the spectral space and are shock-resolved}{For all the simulations presented in this work, we consider $3072\times 3072$ Fourier modes ensuring shock-resolved DNS (see Section~\ref{sec: simulation_parameters} and~\ref{sec: length_scales} for details)}. \Revision{We confirm the shock-resolved nature of the simulations from binned spectral energy, spectral energy transfer function, and spectral dissipation}{Smooth variation of binned spectral energy, spectral energy transfer function, and cumulative spectral dissipation at higher wavenumbers also highlight the shock-resolved nature of the simulations}. For all simulations, we consider constant viscosity ($\mu = 1$) and thermal conductivity ($\alpha = 1$). We show in Appendix~\ref{app: hyper} that temperature dependent viscosity \Revision{}{and thermal conductivity} \Revision{has}{have} no significant affect on shock wave field. Since the shock waves in our simulations are weak, temperature fluctuations and hence viscosity fluctuations are very small compared to their respective base state values to have any significant impact. Furthermore, for simplicity, we ignore bulk viscosity in our simulations ($\kappa = 0$).

At time $t=0$, we prescribe quiescent conditions $\bfu=0$ along with isobaric thermal gradients. To generate a thermal gradient at a wavevector $\bfk$ randomly, we define the initial temperature field $1/f(x,y)$ in Eq.~\eqref{eq: thermal_gradient_init} as, 
\begin{equation}
\frac{1}{f(x,y)}= 1 + \epsilon_T T^b(x,y),
\end{equation}
where $T^b(x,y)$ is defined as a random periodic field for a wavenumber vector $\bfk$ in the Fourier space as,
\begin{equation}
\widehat{T}^b_{\bfk} = \exp(2\pi i\theta_{\bfk}),
\end{equation}
where $\theta_{\bfk}$ is drawn from a normal distribution over $(0,1]$ for every wavenumber vector $\bfk$. Since these thermal gradients are assumed to be isobaric, the initial density field is set to $f(x,y)$ ensuring no pressure perturbations due to thermal initialization. To study the interaction of these thermal gradients with random shock waves, we consider two types of forcings namely random forcing and energy rescaling as discussed below.

\subsection{Random forcing}
\label{sec: RandomForcing}
\begin{figure*}
\includegraphics[width=1.0\textwidth]{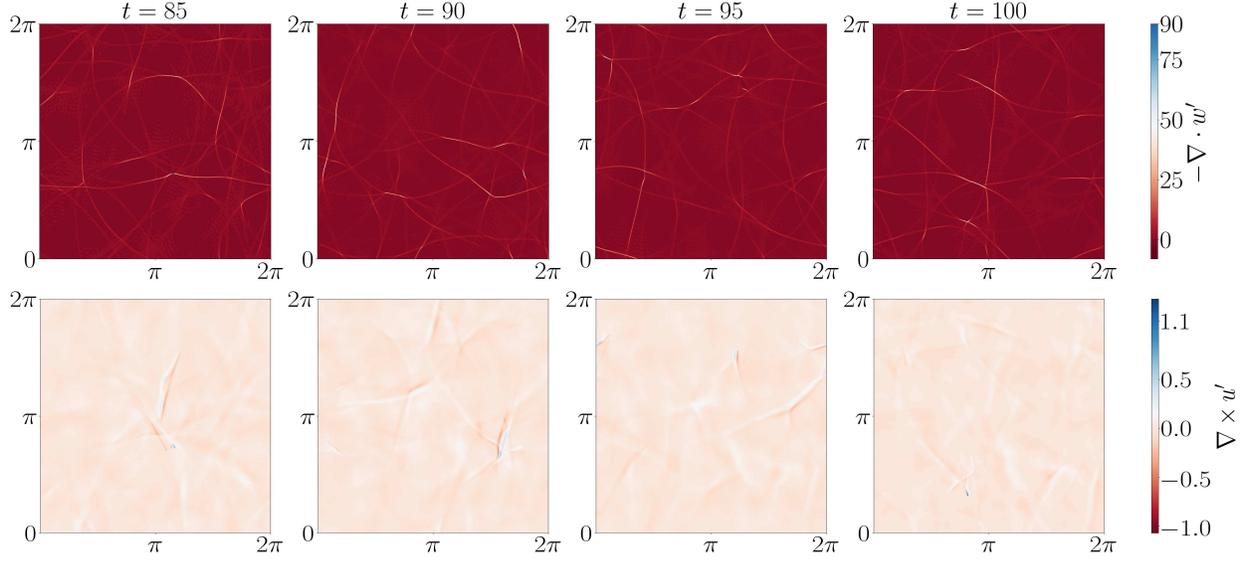}
\put(-520,230){(a)}
\put(-520,115){(b)}
\caption{\Revision{}{(a) Instantaneous contours of $-\nabla\cdot\bfw'$ showing the two-dimensional field of shock waves generated by stochastic forcing and (b) instantaneous contours of $\nabla\times\bfu'$ showing the vorticity field generated due to these shock waves in a inhomogeneous medium (for case 6d in Table~\ref{tab: parameterSpace}) at different times.}}
\label{fig: divVortContours}
\end{figure*}
To generate random shocks, we force acoustic waves stochastically using the Uhlenbeck-Ornstein (UO) process \cite{eswaran1988examination}, typically used for generating homogeneous isotropic turbulence in a three-dimensional box. The forcing $\bfF$ in Eq.~\eqref{eq: momentum_equation} is defined in the Fourier space using four independent solutions of a UO process $a_{i,\bfk}(t)$ for $i=1, 2, 3, $ and $4$ (one complex number in each direction). Following conditions are imposed on each of the $a_{i}$,
\begin{align}
 &\left\langle a_{i, \bfk}(t)\right\rangle = 0 ,\\
  &\left\langle {a}_{i, \bfk}(t) {a}_{j, \bfk}^{*}(t+s)\right\rangle = 2 \sigma^{2} \delta_{ij}\exp(-s/T_L) ,
\end{align}
where $\left\langle\cdot\right\rangle$ denotes the ensemble average, $()^*$ denotes the complex conjugate, $\sigma^2$ is the variance, and $T_L$ is the forcing time-scale. We restrict $\bfF$ to a band of wavenumbers $0<|\bfk|<k_F$. Spectral wave energy cascades from the forcing band to the higher wavenumbers generating shock waves~(see~Fig.~\ref{fig: divVortContours}). In this work, we choose $k_F = 5$ \Revision{}{ensuring energy injection in large length scales~\cite{yeung2018effects, bell2022thermal}}. The total rate of energy addition $\epsilon$ in the system depends on the variance $\sigma^2$, time-scale $T_L$, and number of wavenumber vectors within the forcing band $N_F$ as, 
\begin{align}
  \epsilon = 4N_F T_L \sigma^{2}.
  \label{eq: energyAddition}
\end{align}
Combining independent processes $a_1, a_2, a_3$, and $a_4$, the forcing vector in the Fourier space can be obtained as, 
\begin{equation}
    \widehat{\bfa}_{\bfk} = \left(a_{1,\bfk} + i a_{2,\bfk}, a_{3,\bfk} + ia_{4,\bfk}\right)^T.
    \label{eq: randomForcingVector}
\end{equation}

\begin{figure}[!b]
        \centering
        \includegraphics[width=0.5\textwidth]{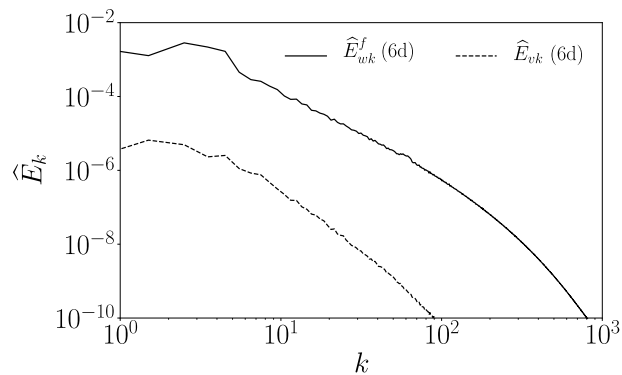}
        \caption{Wave energy $\widehat{E}_{wk}$, and actual vortical energy $\widehat{E}_{vk}$ spectra of stochastically forced simulations for case 6d in Table~\ref{tab: parameterSpace}. Stochastic forcing generates broadband vorticity with highest actual vortical energy in forced wavenumbers due to continuously changing curvature of shock waves.}
        \label{fig: RandomBroadband}
    \end{figure}

To force only the wave field, we remove any vortical component from $\bfa_{\bfk}$ yielding the forcing $\bfF$ introduced in Eq.~\eqref{eq: momentum_equation} in the Fourier space as,
\begin{align}
 \widehat{\bfF}_{\bfk}(t) = \frac{(\widehat{\bfa}_{\bfk}(t) \cdot\bfk)}{|\bfk|^2}\bfk.
\end{align}
Consequently, Eq.~\eqref{eq: energyAddition} yields an upper bound on the rate of energy addition rather than holding exactly.
Even though we remove the vortical component from the forcing, randomly forced acoustic wave turbulence exhibits broadband vortical energy (see Fig.~\ref{fig: RandomBroadband}) because the forcing $\bfF$ (chosen from a stochastic process) may point in a different direction at different times. Consequently, acoustic density gradient $\nabla \rho(t_1)$ generated by $\bfF(t_1)$ can interact with $\bfF(t_2)$ at time $t_2>t_1$, when $\bfF(t_2)$ is not parallel to $\nabla \rho (t_1)$, thus changing the curvature in the shocks generating vorticity. Figure~\ref{fig: divVortContours} shows the two-dimensional field of random shock waves generated due to the stochastic forcing and the vorticity generated by the shock waves interacting with the background thermal gradients ($k_T=60$) and the forcing vector $\bfF$.

\subsection{Density weighted energy rescaling}
\label{sec: rescaling}
Since the forcing parameter $\widehat{\bfa}_{\bfk}$ in Eq.~\eqref{eq: randomForcingVector} is a solution of a Langevin equation, a randomly forced wave turbulence can be compared to the Langevin thermostat in the molecular simulations~\cite{j2007statistical}. At every time-step, components of $\widehat{\bfa}_k$ satisfy the Langevin equation independently. Hence, phase of the acoustic waves generated by stochastic forcing is random. Consequently, vorticity generated as a result of the baroclinic torque due to shock waves propagating in one direction is quickly nullified by shock waves propagating in opposite direction at some later time. Hence, vorticity is not accumulated in the domain at any particular length scale due to baroclinic torque. However, due to interaction of the forcing vector $\bfF$ with the shock waves, a broadband vorticity with higher energy at forced length scales is generated.

Based on this reasoning, we use an \emph{energy rescaling} forcing, which is inspired by the velocity rescaling thermostat in molecular simulations~\cite{j2007statistical} and the deterministic forcing used in hydrodynamic turbulence~\cite{ghosal1995dynamic}. For one set of simulations with different length scales of background thermal gradients, we restart the simulations by disabling random forcing $\bfF$ after statistical stationarity is achieved approximately. In simulations with energy rescaling, density-weighted wave spectral energy $\widehat{E}^f_{wk}$ is rescaled in a band of wavenumbers $\mathcal{K}$ by forcing the contribution of smaller bins of width $\Delta k$ in $\mathcal{K}$ uniform. For instance, let a band of wavenumbers $\mathcal{K}$ defined as $\mathcal{K} = \{\bfk : k_1 < |\bfk| < k_2\}$ be composed of partitions $\mathcal{K}_i = \{\bfk : k_i - 1/2 < |\bfk| < k_i + 1/2\}$ where $k_i$ is between $k_1$ and $k_2$. We rescale the variables $\widehat{\bfw}_{w\mathcal{K}_i}$ and $\widehat{p}_{\mathcal{K}_i}$ to $\tilde{\widehat{\bfw}}_{w\mathcal{K}_i}$ and $\tilde{\widehat{p}}_{\mathcal{K}_i}$, respectively, 
\begin{align}
 &\tilde{\widehat{\bfw}}_{w\mathcal{K}_i} = \epsilon_A\frac{\alpha_{k_i}\widehat{\bfw}_{w\mathcal{K}_i}}{\sqrt{\frac{1}{2}\sum_{j}\alpha^2_{k_j}\left(|\widehat{\bfw}_{w\mathcal{K}_j}|^2 + |\widehat{p}_{\mathcal{K}_j}|^2\right)}}, \label{eq: rescaling_variables_1}\\
 &\tilde{\widehat{p}}_{\mathcal{K}_i} = \epsilon_A\frac{\alpha_{k_i}\widehat{p}_{\mathcal{K}_i}}{\sqrt{\frac{1}{2}\sum_{j}\alpha^2_{k_j}\left(|\widehat{\bfw}_{w\mathcal{K}_j}|^2 + |\widehat{p}_{w\mathcal{K}_j}|^2\right)}},
 \label{eq: rescaling_variables_2}
\end{align}
where the weight coefficients $\alpha_{k_i}$ are chosen as, 
\begin{equation}
 \alpha_{k_i} = \frac{1}{\sqrt{\frac{N_{i}}{2}\sum_{\mathcal{K}_i} \left(|\widehat{\bfw}_{w\mathcal{K}_i}|^2 + |\widehat{p}_{\mathcal{K}_i}|^2\right) }},
 \label{eq: rescaling weights}
\end{equation}
where $N_i$ is the number of wavenumber vectors $\bfk$ within the partition $\mathcal{K}_i$.
Thus, all the partitions $\mathcal{K}_i$ contribute equally to the total wave energy in $\mathcal{K}$. The total wave energy in $\mathcal{K}$ is equal to $\epsilon^2_A$ after every rescaling operation. 
Rescaling of wave components of $\bfw'$ in the Fourier space results in rescaling of the amplitudes in the real space, maintaining the propagation direction of the waves at each time step. Furthermore, since we are forcing the density-weighted wave components $\bfw'_w$, energy rescaling also results in external injection of vorticity. However, as we show in Section~\ref{sec: Enstrophy Budget}, the baroclinic interaction between the shock waves and the background thermal gradients are also significant thus resulting in \emph{coherent} vorticity, unlike random forcing which results in broadband vorticity.

\subsection{Simulation parameters}
\label{sec: simulation_parameters}
\begin{figure}
 \centering
 \includegraphics[width=0.5\textwidth]{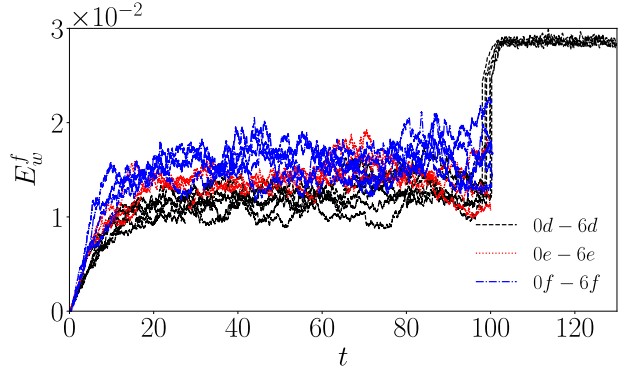}
 \caption{\Revision{}{Time series of $E^f_{w}$ for all the simulation cases in Table~\ref{tab: parameterSpace} and the corresponding baseline cases in Table~\ref{tab: baselineCases}. All simulations are run with stochastic forcing till $t=100$. Only simulations 1d - 6d are restarted with the density-weighted energy rescaling (marked with $^r$ in Table~\ref{tab: parameterSpace}). Due to forcing, spatially averaged pressure and temperature rise in the domain. The shown time series are obtained using pressure perturbations which are calculated by subtracting the spatial average of pressure from total pressure.}}
 \label{fig: TimeSeriesAll}
\end{figure}

We define our simulation parameter space as the wavenumber of thermal perturbations $k_T$ and the rate of energy addition $\varepsilon$. We choose $k_T$ values between $10$ and $60$ (increments of $10$) and $\varepsilon/4N_f$ values as $1.5\times 10^{-5}$, $2.0\times 10^{-5}$,~and $2.5\times 10^{-5}$ (see Table~\ref{tab: parameterSpace}). We run these simulations till $t=100$ and then restart with energy rescaling enabled and stochastic forcing disabled to isolate the coherent vorticity generated due to the interaction between the shock waves and background thermal gradients. Since the parameter $\varepsilon$ does not influence rescaled runs, we only restart simulations with $\varepsilon = 1.5\times 10^{-5}$ with energy rescaling (1d$^r$-6d$^r$ in Table~\ref{tab: parameterSpace}). For each $\varepsilon$ value, we also run a baseline case with no thermal gradients ($f=1$). Figure~\ref{fig: TimeSeriesAll} shows time series of density-weighted wave energy ${E}^f_w$ for all simulations in Tables~\ref{tab: parameterSpace} and~\ref{tab: baselineCases}. \Revision{We note that for}{For} simulations with stochastic forcing, spatially averaged pressure and temperature rise due to constant dissipation of energy caused by shock waves in the system. In Fig.~\ref{fig: TimeSeriesAll}, spatially averaged pressure is removed from $E^f_w$, thus showing the true perturbation energy in the system. For all cases with random gradients, thermal gradients are defined for a range of wavenumber vectors  $k_T - \frac{1}{2} < |\bfk| < k_T + \frac{1}{2}$. 
\begin{table}[!t]
\def\arraystretch{1.5}
\setlength{\tabcolsep}{5.5pt}\
\caption{Simulation parameter space considered in this study and the corresponding simulation names. We run all the simulations using random forcing till $t=100$ and then restart the simulation cases marked with $^r$ with energy rescaling to isolate the coherent baroclinic vorticity. \Revision{We note that the}{The rate of energy injection parameter} $\varepsilon$ loses its significance in rescaled forcing simulations since there is no random forcing. Other fixed parameters and their values are : $\varepsilon_A = 0.15,~\varepsilon_T = 0.1,~Re_{ac} = 2500,$ and$~\mathrm{Pr}=0.72,~$. }
\begin{tabular}{|l|c|cccccc|}
\hline
\multirow{2}{*}{}                & \multicolumn{1}{l|}{\multirow{2}{*}{}} & \multicolumn{6}{c|}{$k_T$}                                                                                                                                \\ \cline{3-8} 
                                 & \multicolumn{1}{l|}{}                  & \multicolumn{1}{l|}{10} & \multicolumn{1}{l|}{20} & \multicolumn{1}{l|}{30} & \multicolumn{1}{l|}{40} & \multicolumn{1}{l|}{50} & \multicolumn{1}{l|}{60} \\ \hline
\multirow{3}{*}{$\epsilon/4N_F$} & $1.5\times 10^{-5}$                    & \multicolumn{1}{c|}{1d$^r$} & \multicolumn{1}{c|}{2d$^r$} & \multicolumn{1}{c|}{3d$^r$} & \multicolumn{1}{c|}{4d$^r$} & \multicolumn{1}{c|}{5d$^r$} & 6d$^r$                      \\ \cline{2-8} 
                                 & $2.0\times 10^{-5}$                    & \multicolumn{1}{c|}{1e} & \multicolumn{1}{c|}{2e} & \multicolumn{1}{c|}{3e} & \multicolumn{1}{c|}{4e} & \multicolumn{1}{c|}{5e} & 6e                      \\ \cline{2-8} 
                                 & $2.5\times 10^{-5}$                    & \multicolumn{1}{c|}{1f} & \multicolumn{1}{c|}{2f} & \multicolumn{1}{c|}{3f} & \multicolumn{1}{c|}{4f} & \multicolumn{1}{c|}{5f} & 6f                      \\ \hline
\end{tabular}
\label{tab: parameterSpace}
\end{table}
\begin{table}[!t]
\def\arraystretch{1.5}
\setlength{\tabcolsep}{5.5pt}
\caption{Baseline simulation cases for $f=1$. }
\begin{tabular}{|c|c|c|c|c|c|c|}
\hline
$\epsilon/4N_F \times 10^{5}$ & $0.25$ & $0.5$ & $1.0$ & 1.5 & 2.0 & 2.5 \\ \hline
                              & 0a     & 0b    & 0c    & 0d  & 0e  & 0f  \\ \hline
\end{tabular}
\label{tab: baselineCases}
\end{table}

\begin{figure}[!t]
        \centering
        \includegraphics[width=0.5\textwidth]{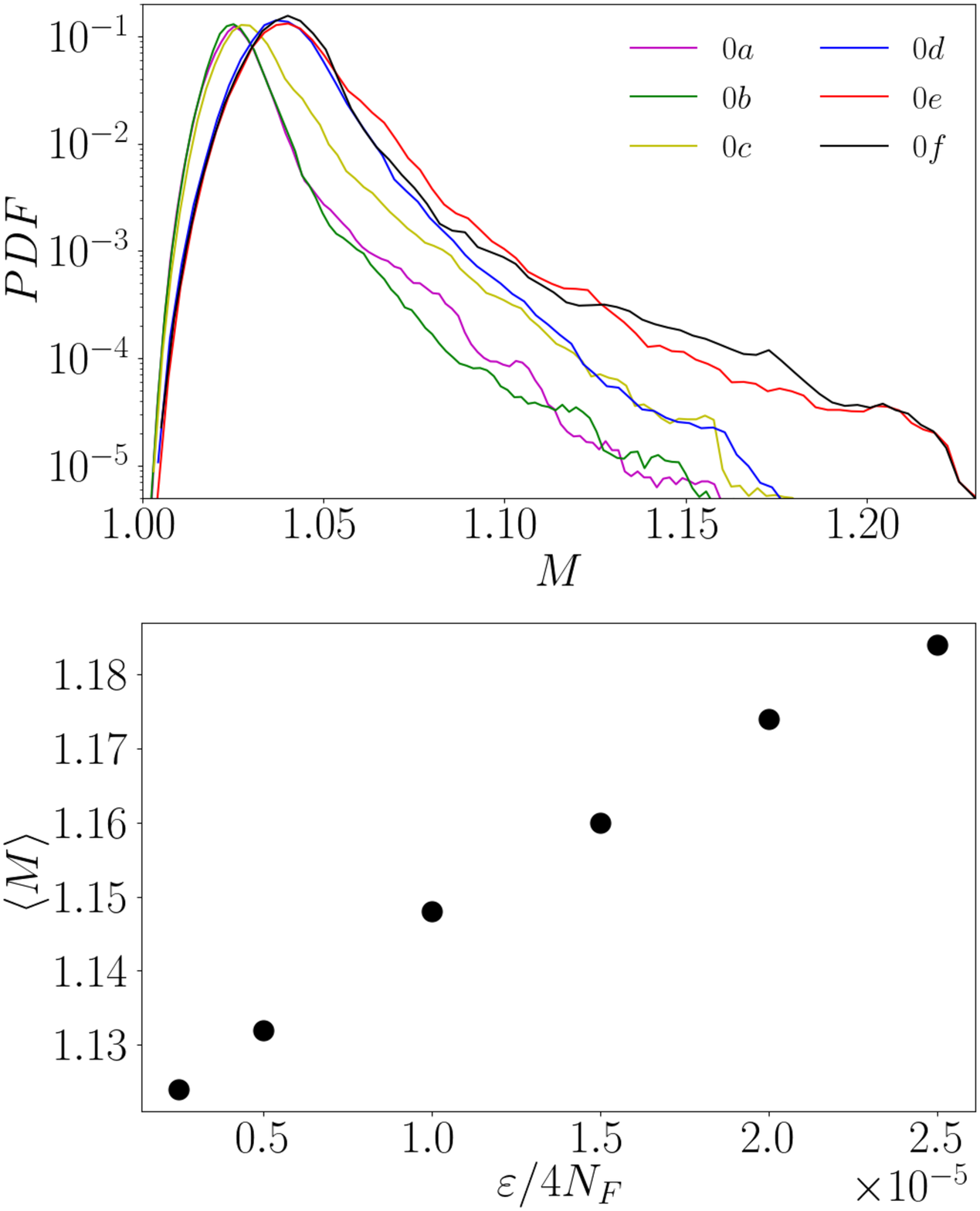}
        \put(-255,315){(a)}
        \put(-255,150){(b)}
        \caption{\Revision{}{(a) Time averaged normalized histogram (PDF) of Mach number, (b) variation of weighted mean Mach number $\langle M\rangle$ with forcing parameter $\epsilon$ for baseline cases 0a-0f (see Table~\ref{tab: baselineCases}).}}
        \label{fig: MachEps}
    \end{figure}

The acoustic Reynolds number $Re_{ac}$ and the rate of energy \Revision{dissipation}{injection} $\epsilon$ (equal to rate of energy \Revision{addition}{dissipation} at stationarity) are related as~\cite{gupta2018spectral, augier2019shallow}, 
\begin{equation}
Re_{ac}  \sim \frac{1}{\eta } (\epsilon \ell)^{-1/3} ,
     \label{eq: Kolomogrov scale}    
\end{equation}
where $\eta$ is the smallest length scale (Kolmogorov length scale in hydrodynamic turbulence or shock thickness) and $\ell$ is the integral length scale (average distance between two shocks). \Revision{}{Since 3072 Fourier modes are used in each direction with $2/3$ dealiasing, maximum wavenumber captured in the simulations is $k_{max}=1024$. Hence, for shock-resolved simulations, $k_{max}\eta > 1$ must hold (see Section~\ref{sec: length_scales}).} We choose $Re_{ac} = 2500$ in our simulations such that $\eta > 1/1024$ for all the values of $\varepsilon$ and the field of shock waves is well resolved \Revision{}{(see Table~\ref{tab: Time averaged Length scales})}. \RevisionTwo{}{For all the values of $\varepsilon$, the shock waves in the field are weak shock waves~(see Fig.~\ref{fig: MachEps}a).}

\Revision{}{As the rate of energy injection $\epsilon$ increases, stronger shocks are generated in the domain. We calculate the mean Mach number $\langle M \rangle$ using the entropy perturbation at each point in the simulation domain and calculating the weighted mean of the shock Mach number. At each point, we evaluate the dimensionless entropy perturbation $s'$ using, }
\begin{equation}
 s' = \frac{1}{\gamma - 1}\ln\left(\gamma p'\right) - \frac{\gamma }{\gamma - 1}\ln\left(\frac{\rho'}{f}\right),
\end{equation}
using which, we approximate the entropy jump at each point as, 
\begin{equation}
 \Delta s' = \Bigg|\frac{\partial s'}{\partial x}\Bigg|dx + \Bigg|\frac{\partial s'}{\partial y}\Bigg|dy. 
\end{equation}
Using $\Delta s'$, we calculate the Mach number field assuming weak shocks as, 
\begin{equation}
 M = \sqrt{1 + \left(\frac{3\Delta s'(\gamma + 1)^2}{2\gamma}\right)^{1/3}}.
\end{equation}
\Revision{}{In Fig.~\ref{fig: MachEps} we show the normalized histogram or probability distribution function (PDF) of Mach number $M$ for all the baseline simulations. Since the shock waves are very thin, most of the domain exhibits Mach number very close to 1 ($s'\approx 0$). To obtain a single numerical value representative of the Mach number of shock waves, we compute weighted average of the Mach number $\langle M \rangle$, such that contribution of points where $s' < 25\%$ of the maximum entropy perturbation is set to 0. Figure~\ref{fig: MachEps}(b) shows the variation of weighted mean Mach number $\langle M\rangle$ of shock waves in the domain with the forcing energy injection rate $\epsilon$.} 
Since the shocks become stronger for higher energy injection rate for same forcing wavenumber band, the smallest length scale of fields decreases (see Eq.~\eqref{eq: Kolomogrov scale}) and the mean Mach number of shocks increases. \Revision{We note that even}{Even} for the highest injection rate considered in this study \Revision{}{(case 0f) the shocks are weak ($\langle M \rangle < 1.2$)}. Thus the governing equations correct up to second order in perturbations (Eqs.~\eqref{eq: momentum_nonlinear}-\eqref{eq: pressure_nonlinear}) are sufficient to study the cascade of spectral wave energy. To elucidate the generation of vorticity, we consider the fully nonlinear Eq.~\eqref{eq: momentum_equation} (see Section~\ref{sec: Vorticity}).
Below, we discuss the features of field of shock waves generated using stochastic forcing and the rescaled forcing in detail. We study the length scales of the field of shock waves following which we focus on vorticity generated due to the two forcing schemes and finally discuss the enstrophy budgets.

\section{Results and Discussion}
\label{sec: resultsDiscussion}
In this section, we discuss the interaction of stochastically forced shocks and shocks maintained by energy rescaling with the background thermal gradients. We show that broadband vorticity is generated due to stochastically forced shocks \Revision{even in the presence of background thermal gradients}{irrespective of the presence of the background thermal gradients}. Shocks maintained by energy rescaling interact with the background thermal gradients and generate coherent vorticity, with eddies of the same length scale as that of the thermal gradient. We also show that the background thermal gradients do not significantly affect the length scales of the shocks and spectral energy cascade in shocks.

\subsection{Length scales}
\label{sec: length_scales}
\begin{figure}[!b]
        \centering
        \includegraphics[width=0.5\textwidth]{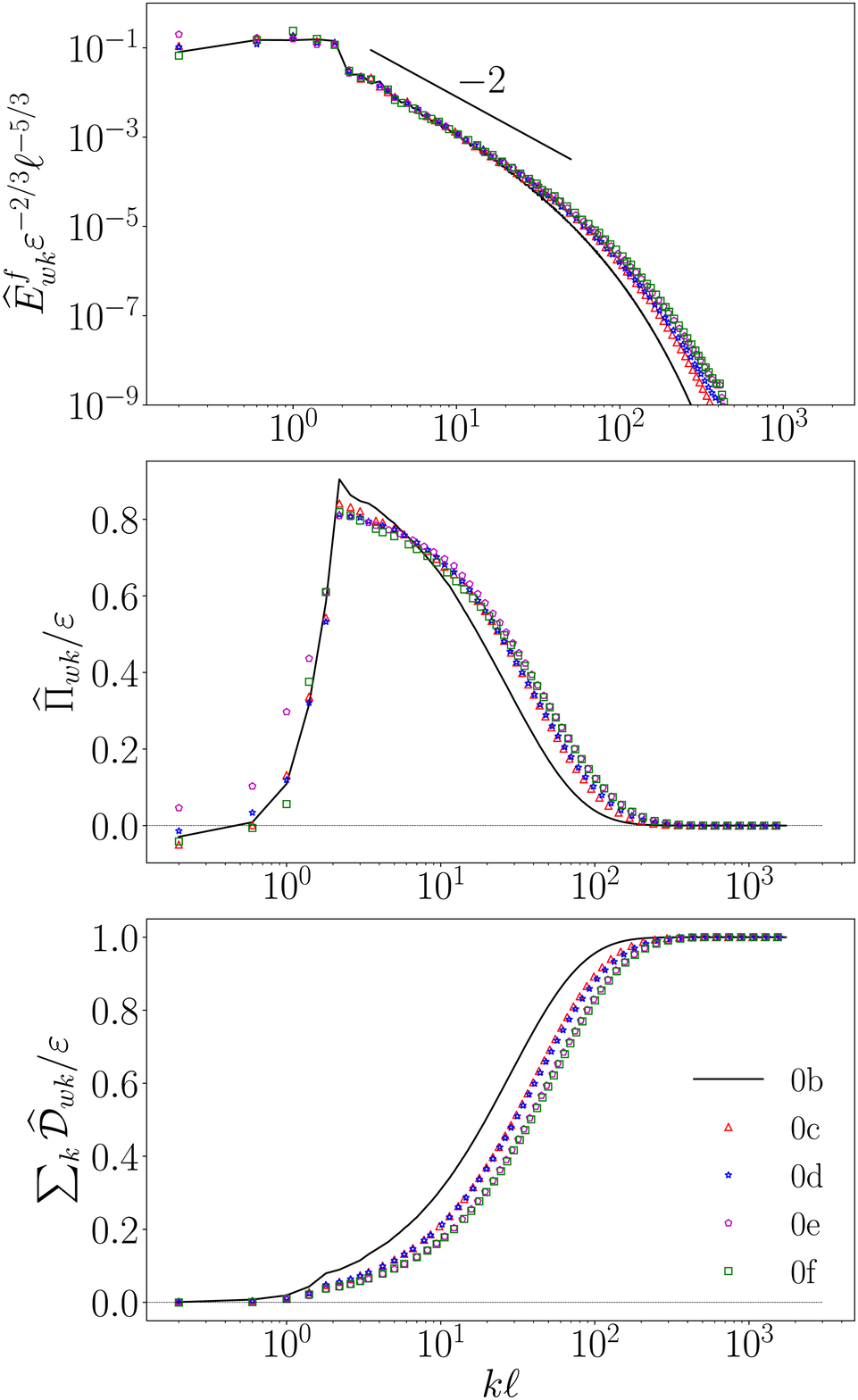}
        \put(-250,415){(a)}
        \put(-250,280){(b)}
        \put(-250,140){(c)}
        \caption{\Revision{}{(a) Dimensionless scaled density-weighted wave energy spectra $\widehat{E}^f_{wk}$, (b) spectral flux of wave energy $\widehat{\Pi}_{wk}$, and (c) cumulative dissipation $\widehat{\mathcal{D}}_{wk}$ for baseline cases (0b-0f in Table~\ref{tab: baselineCases}). The wave energy spectra $\widehat{E}^f_{wk}$ scale with dissipation $\varepsilon$ and integral length scale $\ell$ as in Eq.~\eqref{eq: energySpectraScaling}.}}
        \label{fig: BaselineSpectra}
    \end{figure}
Due to shock waves, the wave energy spectra decay as $k^{-2}$ in the wavenumber range where spectral energy flux is high. \citet{gupta2018spectral} derived such scaling for acoustic wave energy spectra for decaying one-dimensional acoustic wave turbulence in a homogeneous medium as, $\widehat{E}_{wk}\sim \varepsilon^{2/3}\ell^{-1/3}k^{-2}$ ($\widehat{E}^f_{wk} = \widehat{E}_{wk}$ for $f=1$). Augier et al.~\cite{augier2019shallow} derived a similar scaling for stochastically forced isotropic shocks in shallow water equations. In Fig.~\ref{fig: BaselineSpectra} we show the normalized density-weighted wave energy spectra, spectral flux, and cumulative spectral dissipation for baseline cases 0a-0f (see Table~\ref{tab: baselineCases}). \Revision{We note that}{The} wave energy spectra scale as $k^{-2}$ with a similar dependence on dissipation and integral length scale as for decaying one-dimensional acoustic wave turbulence~\cite{gupta2018spectral} and two-dimensional shallow water wave turbulence~\cite{augier2019shallow}, indicating that the scaling relation, 
\begin{equation}
 \frac{\widehat{E}^f_{wk} }{\ell^{-5/3}\epsilon^{2/3}} \sim \frac{1}{k^2\ell^2},
 \label{eq: energySpectraScaling}
\end{equation}
holds where the spectral energy flux is high ($10\lesssim k\ell\lesssim 300$). \RevisionTwo{We note that it}{It} is not possible to term the range of scales where the spectral energy flux is high as \emph{inviscid range} or \emph{inertial range} because the dissipation exists at all scales in shock waves propagating with a physical diffusion term. In Appendix~\ref{app: hyper} we show the effect of hyperdiffusion (widely used in two-dimensional wave turbulence literature) and how it modifies the distribution of spectral energy cascade flux $\widehat{\Pi}_k$ in shock waves, particularly concentrating the dissipation at small scales. Since the wave energy spectra decay as $k^{-2}$ in a field of shock waves, dissipation is active at all length scales. Consequently, cascade of wave energy spectra in compressible flows may be regarded as a variable flux energy transfer~\cite{verma2019energy, verma2021variable}. The scaling in Eq.~\eqref{eq: energySpectraScaling} also indicates that the shocks generated by stochastic forcing in all the simulations are isotropic~\cite{augier2019shallow}.
\begin{figure}[!t]
        \centering
        \includegraphics[width=0.5\textwidth]{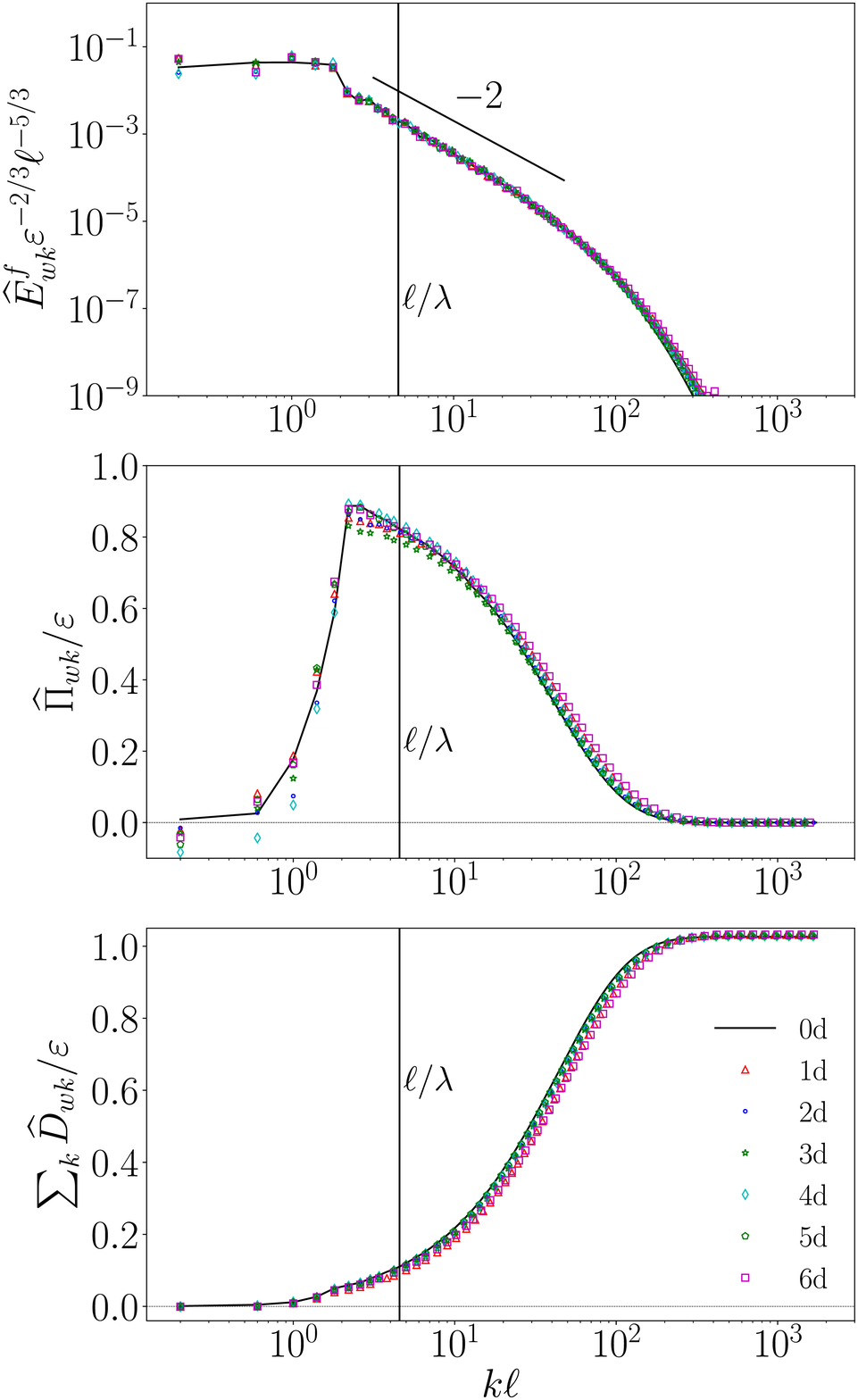}
        \put(-250,415){(a)}
        \put(-250,280){(b)}
        \put(-250,140){(c)}
        \caption{\Revision{}{(a) Dimensionless scaled density-weighted wave energy spectra $\widehat{E}^f_{wk}$, (b) spectral flux of wave energy $\widehat{\Pi}_{wk}$, and (c) cumulative dissipation $\widehat{\mathcal{D}}_{wk}$ for inhomogeneous cases (1d-6d in Table~\ref{tab: parameterSpace} and the baseline case (0d in Table~\ref{tab: baselineCases}). The density-weighted wave energy spectra are almost identical for all the cases and follow the scaling in Eq.~\eqref{eq: energySpectraScaling}. The Taylor microscale $\lambda$ for all the cases is almost identical highlighting that background thermal gradients and length scales of these gradients do not affect the shock waves significantly. }}
        \label{fig: RandomInhomogeneousSpectra}
    \end{figure}
    
The integral length scale $\ell$ quantifies the large length scales at which energy is injected in the system. Since we choose $k_F=5$, we use $\ell = \frac{1}{5/2}$ in Fig.~\ref{fig: BaselineSpectra} for scaling. We identify the Taylor microscale as~\cite{gupta2018spectral}
\begin{equation}
    \lambda \sim \sqrt{\frac{\sum_{k}E_{k}}{\sum_{k} k^{2}E_{k} }}.
    \label{eq: taylorMicroscale}
\end{equation}
Assuming the spectral energy $\widehat{E}^f_{wk}$ constant from $0<k<2/\ell$, we obtain the relation, 
\begin{equation}
 \lambda\sim\sqrt{\eta\ell},
 \label{eq: LengthScales}
\end{equation}
which indicates that the Taylor microscale is in the middle of the integral length scale and the Kolmogorov length scale, similar to one-dimensional acoustic wave turbulence and unlike classical hydrodynamic turbulence in which $\lambda\sim \eta^{2/3}\ell^{1/3}$ and is biased towards the smallest length scale indicating that the dissipation is higher at smaller length scales~\cite{pope2000turbulent}. Hence, scaling in Eq.~\eqref{eq: LengthScales} also shows that the dissipation occurs uniformly across all length scales smaller than the integral length scale, particularly in the range of wavenumbers for which $\widehat{\Pi}_{wk}$ is high. Figure~\ref{fig: RandomInhomogeneousSpectra} shows the wave energy spectra  $\widehat{E}^f_{wk}$, flux of spectral wave energy $\widehat{\Pi}_{wk}$, and cumulative dissipation of spectral wave energy $\widehat{\mathcal{D}}_{wk}$ for simulation cases 0d-6d. Background thermal gradients do not affect the wave energy spectra and the length scales associated with the cascade of energy. \Revision{}{Table~\ref{tab: Time averaged Length scales} shows the time-averaged values of the Taylor microscale $\langle\lambda\rangle_t$ calculated using Eq.~\eqref{eq: taylorMicroscale} and the Kolmogorov length scale $\langle\eta\rangle_t$ calculated using Eq.~\eqref{eq: Kolomogrov scale} for cases 0d-6d (see Table~\ref{tab: parameterSpace}). Both the length scales change insignificantly as the length scales of thermal gradients change. Moreover, for all cases $k_{max} \eta$ remains greater than 1 confirming the shock-resolved nature of simulations.} As we discuss further, baroclinic interactions with background thermal gradients result in change of vortical length scales generated in the domain.
  \begin{table}[!t]
    \centering
\def\arraystretch{1.5}
\setlength{\tabcolsep}{3.25pt}
\caption{\Revision{}{Time-averaged Taylor microscale $\langle \lambda \rangle_t$ and Kolmogorov scale $\langle \eta \rangle_t $. }}
\begin{tabular}{|c|c|c|c|c|c|c|c|}
\hline
Case & 0d & 1d & 2d & 3d & 4d & 5d & 6d \\ \hline
                              $\langle \lambda \rangle_t $& 0.087     & 0.088   & 0.083    & 0.089  & 0.084  & 0.086 & 0.088  \\ \hline
                              $ \langle \eta \rangle_t $& 0.0030     & 0.0027   & 0.0028    & 0.0028  & 0.0028  & 0.0030 & 0.0028  \\ \hline
\end{tabular}
\label{tab: Time averaged Length scales}
\end{table}
\subsection{Vorticity}
\label{sec: Vorticity}
Due to spectral energy cascade, pressure gradients exist at a wide range of length scales (from largest length scale $\ell$ to the smallest length scale $\eta$). These pressure gradients can interact with the density gradients in the domain to generate local baroclinic rotation of fluid, quantified by vorticity. \Revision{Based on}{Motivated by} the decomposition in Eqs.~\eqref{eq: Decomposition}, we identify vorticity $\bfOmega$ and density-weighted vorticity  $\bfOmega_{f}$ as, 
\begin{equation}
 \bfOmega = \nabla\times\bfu,~~\bfOmega_{f} = \nabla\times\left(\sqrt{\rho}\bfu\right).
\end{equation}
Taking curl of Eq.~\eqref{eq: momentum_equation} (with constant $\mu$ and $\kappa = 0$), we obtain the governing equation for $\bfOmega$ (in 2D) as, 
\begin{align}
 \frac{\partial  \bfOmega }{\partial t}&=
 \frac{\nabla \rho \times \nabla p}{\rho^2} +  \frac{\mu\nabla^{2} \bfOmega}{\rho Re_{ac}}  +\frac{\mu}{Re_{ac}} \frac{\nabla \rho \times [ \nabla \times \bfOmega ] }{\rho^2}\nonumber\\
   &-  \left (\frac{4 \mu}{3Re_{ac}} \right) \frac{\nabla \rho \times \nabla (\nabla \cdot \bfu ) }{\rho^{2}}  -\nabla\cdot(\bfu\bfOmega) + \bfF^r_\Omega.
\label{eq: omega}
\end{align}
\begin{figure}[!b]
        \centering
        \includegraphics[width=0.5\textwidth]{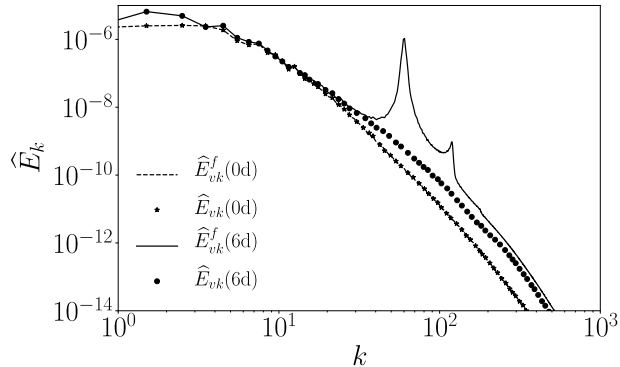}
        \caption{Vortical energy $\widehat{E}_{vk}$ and density-weighted vortical energy $\widehat{E}^f_{vk}$ for homogeneous (baseline) simulation 0d (see Table~\ref{tab: baselineCases}) and simulation with thermal gradients 6d (see Table~\ref{tab: parameterSpace}) with stochastically forced acoustic waves. While $\widehat{E}^f_{vk}$ peaks near $k_T=60$ for 6d, $\widehat{E}_{vk}$ exhibits no such peak, thus highlighting that the rotational energy $\widehat{E}_{vk}$ is broadband when shocks are forced stochastically. }
        \label{fig: EfvVSEvRandom}
    \end{figure}
    \begin{figure}[!t]
        \centering
        \includegraphics[width=0.49\textwidth]{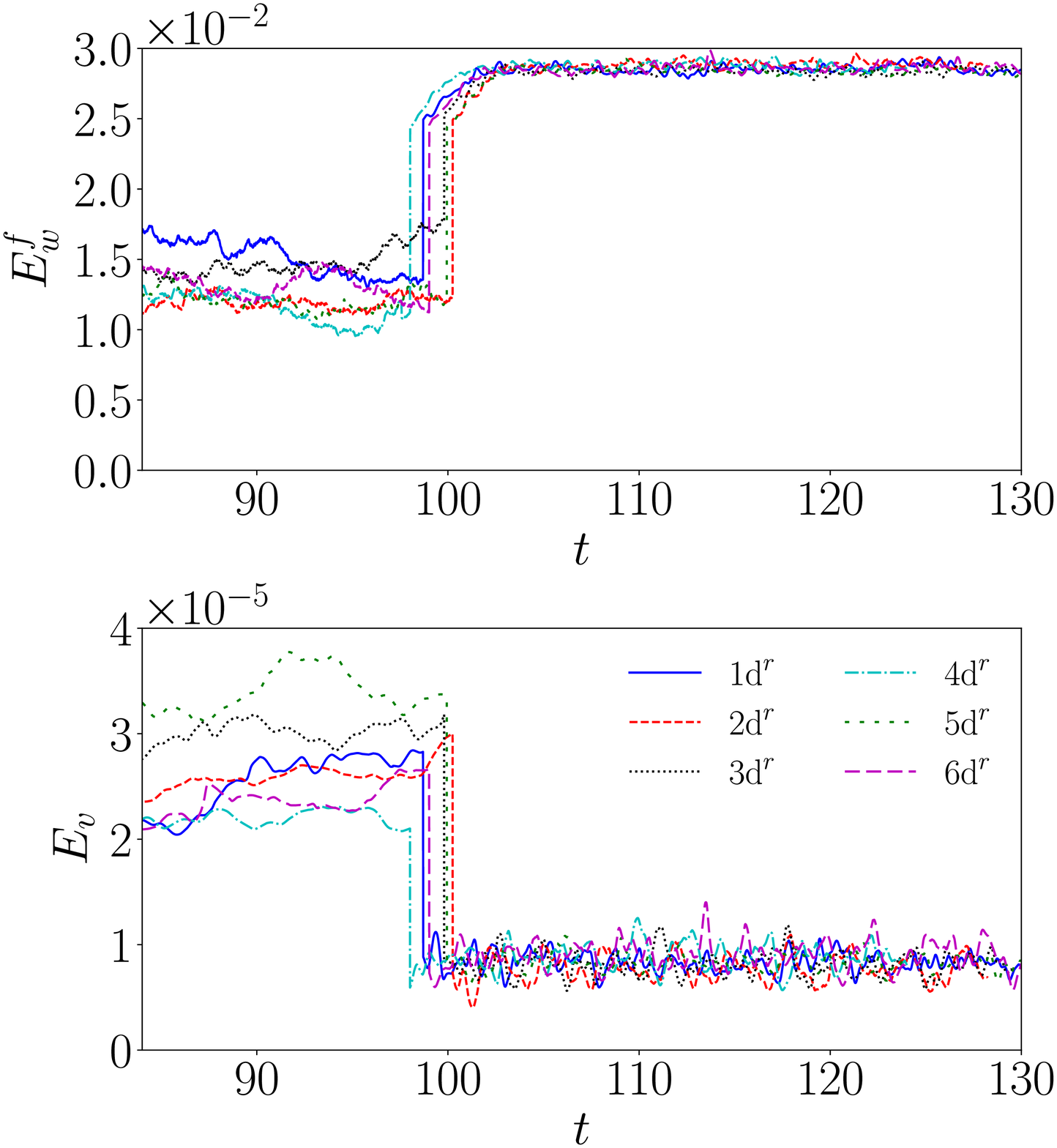}
        \put(-250,265){(a)}
        \put(-250,130){(b)}
        \caption{\Revision{}{Time series of (a) density-weighted wave energy $E^f_w$ and (b) Vortical energy $E_v$ for simulations 1d$^r$-6d$^r$ before and after energy rescaling is enabled. Overall vortical energy in the system decreases and the density-weighted wave energy quickly saturates near $\epsilon^2_A$.}}
        \label{fig: TimeSeriesEwEv}
    \end{figure}

\begin{figure*}
\includegraphics[width=1.0\textwidth]
{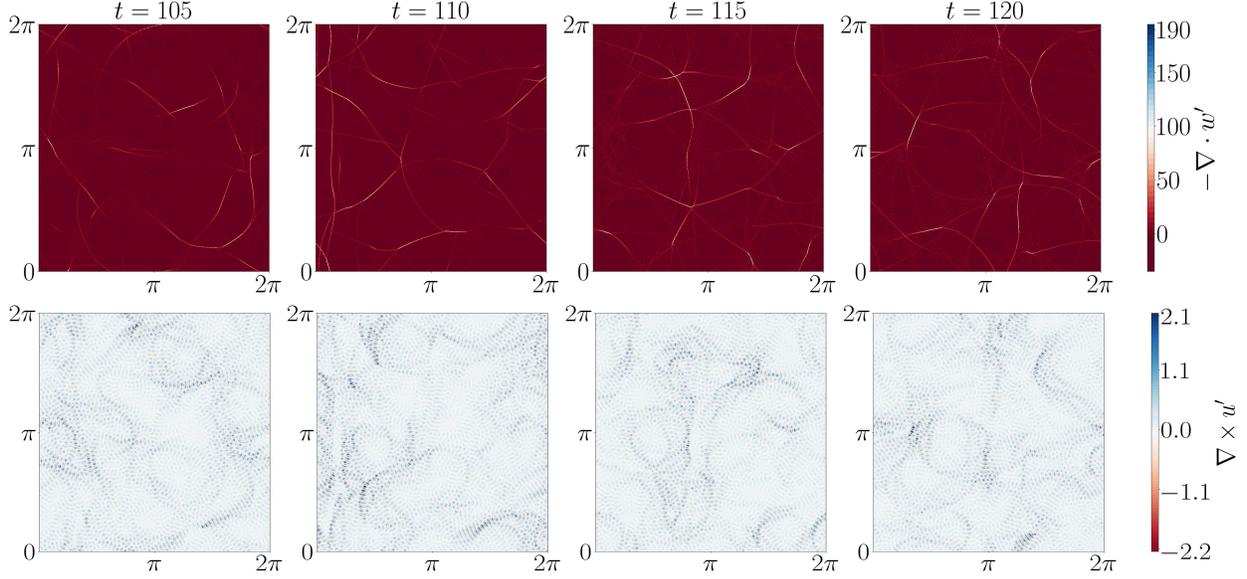}
\put(-520,235){(a)}
\put(-520,115){(b)}
 \caption{\Revision{}{(a) Instantaneous contours of $-\nabla\cdot\bfw'$ showing the two-dimensional field of shock waves generated by rescaled forcing and (b) instantaneous contours of $\nabla\times\bfu'$ showing the coherent vorticity field generated due to the interaction shock waves with the inhomogeneous background medium (for case 6d in Table~\ref{tab: parameterSpace}) at different times.}}
\label{fig: divVortContoursRescaled}
\end{figure*}

\Revision{We note that Eq.}{Equation}~\eqref{eq: omega} holds without any perturbation decomposition of the field variables. First term on the right hand side of Eq.~\eqref{eq: omega} is the baroclinic term. As we show in the next section, interaction of density gradients with pressure gradients is the primary source of vorticity (and enstrophy) generation in our simulations. \RevisionTwo{We note that these}{These} density gradients include both the background thermal inhomogeneity and the density gradients generated in the shock waves. The second term is the viscous diffusion of vorticity. The next two terms represent the interaction of density gradients with viscous stresses in the domain. In a field of shock waves, \Revision{bulk viscous stresses (proportional to $\nabla\cdot\bfu$)}{bulk viscous stress terms (proportional to $\nabla\left(\nabla\cdot\bfu\right)$)} are significantly higher than shear stresses (proportional to $\nabla\times\Omega$) \Revision{}{because velocity dilatation is orders of magnitude higher than the rotation rate (c.f. Figs.~\ref{fig: divVortContours} and \ref{fig: divVortContoursRescaled})}. Hence, we expect the third term in Eq.~\eqref{eq: omega} to be significantly smaller in magnitude compared to the fourth term. Furthermore, an interaction of density gradients with bulk viscous stresses (captured by the fourth term) does not necessarily denote dissipation of vorticity and could also participate in generation of vorticity, as noted by~\citet{kida1990enstrophy}. The next term represents advection of vorticity. The last term $\bfF^r_\Omega$ denotes any forcing of vorticity due to density-weighted wave energy rescaling. \Revision{We note that}{Since} it is difficult to write the energy rescaling as a right hand side in the momentum equation. Hence, we do not simplify $\bfF^r_\Omega$ further. Since the curl of stochastic forcing $\bfF$ is set to zero, there is no stochastic forcing of vorticity $\bfOmega$. \Revision{}{In the next section, we derive the enstrophy budget terms using Eq.~\eqref{eq: omega}}.
Similarly, a governing equation for $\bfOmega_f = \nabla\times\left(\sqrt{\rho}\bfu\right)$ can be derived as,  
\begin{align}
 \frac{\partial  \bfOmega_f }{\partial t}&=
 \frac{\nabla \sqrt{\rho} \times \nabla p}{\rho} +  \frac{\mu\nabla^{2} \bfOmega}{\sqrt{\rho} Re_{ac}}  +\frac{\mu}{Re_{ac}} \frac{\nabla \sqrt{\rho} \times [ \nabla \times \bfOmega ] }{\rho}\nonumber\\
   &-  \left (\frac{4 \mu}{3Re_{ac}} \right) \frac{\nabla \sqrt{\rho} \times \nabla (\nabla \cdot \bfu ) }{\rho}  -\nabla\times(\sqrt{\rho}\bfu\cdot\nabla\bfu) \nonumber \\ &+ \nabla\sqrt{\rho}\times\bfF.
 \label{eq: omega_f}
\end{align}
    \begin{figure}[!b]
        \centering
        \includegraphics[width=0.5\textwidth]{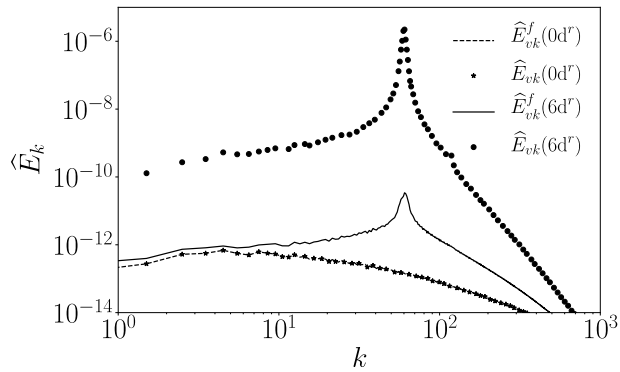}
        \caption{Vortical energy $\widehat{E}_{vk}$ and density-weighted vortical energy $\widehat{E}^f_{vk}$ for homogeneous (baseline) simulation 0d (see Table~\ref{tab: baselineCases}) and simulation with thermal gradients 6d (see Table~\ref{tab: parameterSpace}) with energy rescaling.}
        \label{fig: EfvVSEvRescaled}
    \end{figure}
Since energy rescaling only rescales energy in density-weighted wave modes, density-weighted vorticity $\bfOmega^f$ is not affected by energy rescaling. However, the final term in Eq.~\eqref{eq: omega_f} is due to random forcing interacting with the density gradients. 
Figure~\ref{fig: EfvVSEvRandom} shows the spectra of $\widehat{E}_{vk}$, which quantifies the vorticity $\bfOmega$ in the domain and $\widehat{E}^f_{vk}$, which quantifies the density-weighted vorticity $\bfOmega_f$ for a stochastically forced homogeneous simulation (0d in Table~\ref{tab: baselineCases}) and a stochastically forced inhomogeneous simulation (6d in Table~\ref{tab: parameterSpace}). \RevisionTwo{We note that}{The} random forcing results in generation of a broadband vorticity $\bfOmega$. The peak in $\widehat{E}^f_{vk}$ near the wavenumber $k_T$ occurs due to $\sqrt{f}$ in the definition of $\widehat{E}^f_{vk}$, hence falsely indicating coherent vorticity in the simulation. Consequently, in our further analysis of vorticity in this section and enstrophy budgets in the next section, we consider $\widehat{E}_{vk}$, $\bfOmega$, and the corresponding velocity field $\bfu_v$ (c.f. Eq.~\eqref{eq: rotational_velocity}).

As shown in Fig.~\ref{fig: EfvVSEvRandom}, vorticity is not localized at any particular length scale in both the homogeneous and inhomogeneous simulations. The broadband vorticity is generated due to changing shock curvature by the action of stochastic forcing. This also indicates that if the curvature of the shock waves is maintained, then the interaction with the thermal gradients must generate coherent vorticity at length scale at which thermal gradients are concentrated. To this end, we consider the energy rescaling runs which we restart from the stochastically forced runs at $t=100$. During restart, we remove the vortical velocity $\bfu_v$ (c.f.~Eq.~\eqref{eq: rotational_velocity}) and put $\bfF=0$. During the simulations, the density-weighted spectral energy $\widehat{E}^f_{wk}$ is rescaled as per Eqs.~\eqref{eq: rescaling_variables_1}-~\eqref{eq: rescaling weights} for $\varepsilon_A = 0.15$. 

Figure~\ref{fig: TimeSeriesEwEv} shows the time series of total density-weighted wave energy $E^f_w$ and vortical energy $E_v$ before and after energy rescaling restart. The wave energy readjusts to the energy rescaling and saturates around $\epsilon^2_A$ (see Fig.~\ref{fig: TimeSeriesEwEv}(a)). Since we remove any vortical component $\bfu_v$ from the velocity field, the vortical energy drops but immediately recovers to a small but finite value (approximately $1\times 10^{-5}$ in Fig.~\ref{fig: TimeSeriesEwEv}b) indicating that the vortical velocity field is atleast one order smaller than the wave velocity field. Figure~\ref{fig: EfvVSEvRescaled} shows the spectra of $\widehat{E}_{vk}$ and $\widehat{E}^f_{vk}$ for simulations with energy rescaling in a homogeneous medium (0d$^r$ in Table~\ref{tab: baselineCases}) and a heterogeneous medium with thermal gradients at $k_T=60$ (6d$^r$ in Table~\ref{tab: parameterSpace}). \Revision{We note that the}{The} vortical energy $\widehat{E}^f_{vk}$ is narrow-band and is localized around the length scale corresponding to $k_T$ for the inhomogeneous simulation, indicating coherent vorticity generation. In simulations with energy rescaling, spectral wave energy $\widehat{E}^f_{wk}$ is rescaled (restored) so as to maintain the total wave energy at large length scales constant in time. Recoursing to concepts in hydrodynamic turbulence, energy rescaling ensures \emph{permanence of large eddies} in time. Consequently, the direction in which shock waves are propagating (phasing of shock waves) is never altered by energy rescaling. As a result, there is no broadband vorticity generated in the domain due to changing directions and curvature of shock waves. Instead, a coherent vorticity at the same length scale corresponding to the thermal gradient is generated. Figure~\ref{fig: divVortContoursRescaled}(a) shows the snapshots of two-dimensional field of shock waves generated by the energy rescaling in an inhomogeneous medium with thermal gradients at $k_T=60$ (6d$^r$ in Table~\ref{tab: parameterSpace}). Figure~\ref{fig: divVortContoursRescaled}(b) shows the coherent vorticity generated due to these shock waves.

In the next section, we discuss the budgets of enstrophy defined as $Q = |\bfOmega|^2/2$ indetifying the primary mechanism for generation of coherent vorticity due to shock waves driven by energy rescaling. 
    \begin{figure}[!b]
        \centering
        \includegraphics[width=0.5\textwidth]{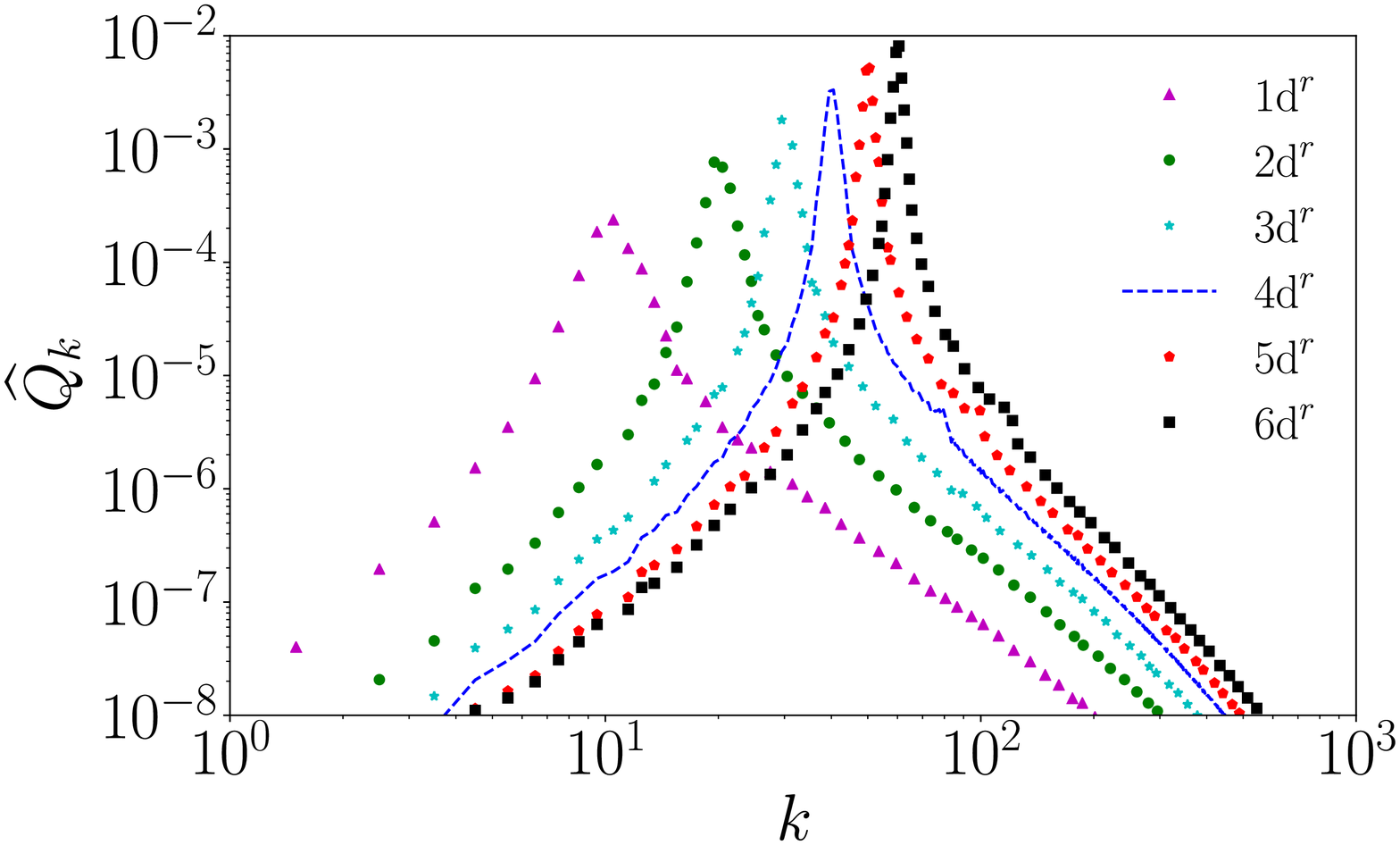}
        \caption{\Revision{}{Enstrophy spectra $\widehat{Q}_k$ for simulations with inhomogeneous thermal gradients 1d$^r$ - 6d$^r$ (see Table~\ref{tab: parameterSpace}) with energy rescaling.}}
        \label{fig: EnstrophySpectra}
    \end{figure}
\subsection{\label{sec: Enstrophy Budget}Enstrophy Budget}
Generation of coherent vorticity is highlighted in enstrophy spectra (see Fig.~\ref{fig: EnstrophySpectra}) defined as,
\begin{equation}
 \widehat{Q}_k = \frac{1}{2}\big|\widehat{\bfOmega}_k\big|^2.
\end{equation}
Using Eq.~\eqref{eq: omega}, we obtain the governing equation for enstrophy spectra as, 
\begin{align}
 \frac{d\widehat{Q}_k}{dt} = \widehat{B}_k - \widehat{D}_k - \widehat{A}_k + \widehat{F}^r_k,
  \label{eq: EnstrophySpectra}
 \end{align}
 where
 \begin{align}
 \widehat{B}_k &= \mathrm{Re}\left[\widehat{\bfOmega}^*_k\cdot\widehat{\left(\frac{\nabla \rho \times \nabla p}{\rho^2}\right)}_k\right],\label{eq: enstrophy_baroclinic}\\
 \widehat{D}_k &= -\frac{\mu}{Re_{ac}}\mathrm{Re}\Bigg[\widehat{\bfOmega}^*_k\cdot\Bigg(\frac{\nabla^2\bfOmega}{\rho} + \frac{\nabla\rho\times\nabla\times\bfOmega}{\rho^2}  \nonumber\\
 &-\frac{4}{3}\frac{\nabla\rho\times\nabla(\nabla\cdot\bfu)}{\rho^2}\Bigg)_k\Bigg],~\label{eq: enstrophy_dissipation}\\
 \widehat{A}_k &= \mathrm{Re}\left[\widehat{\bfOmega}^*_k\cdot\left(\widehat{\nabla\cdot\left(\bfu\bfOmega\right)}\right)_k\right],~\label{eq: enstrophy_adv}
\end{align}
and $\widehat{F}^r_k$ denotes the enstrophy generation due to energy rescaling (term corresponding to $\bfF^r_\Omega$ in Eq.~\eqref{eq: omega}).
\begin{figure}[!t]
\centering
 \includegraphics[width=0.5\textwidth]{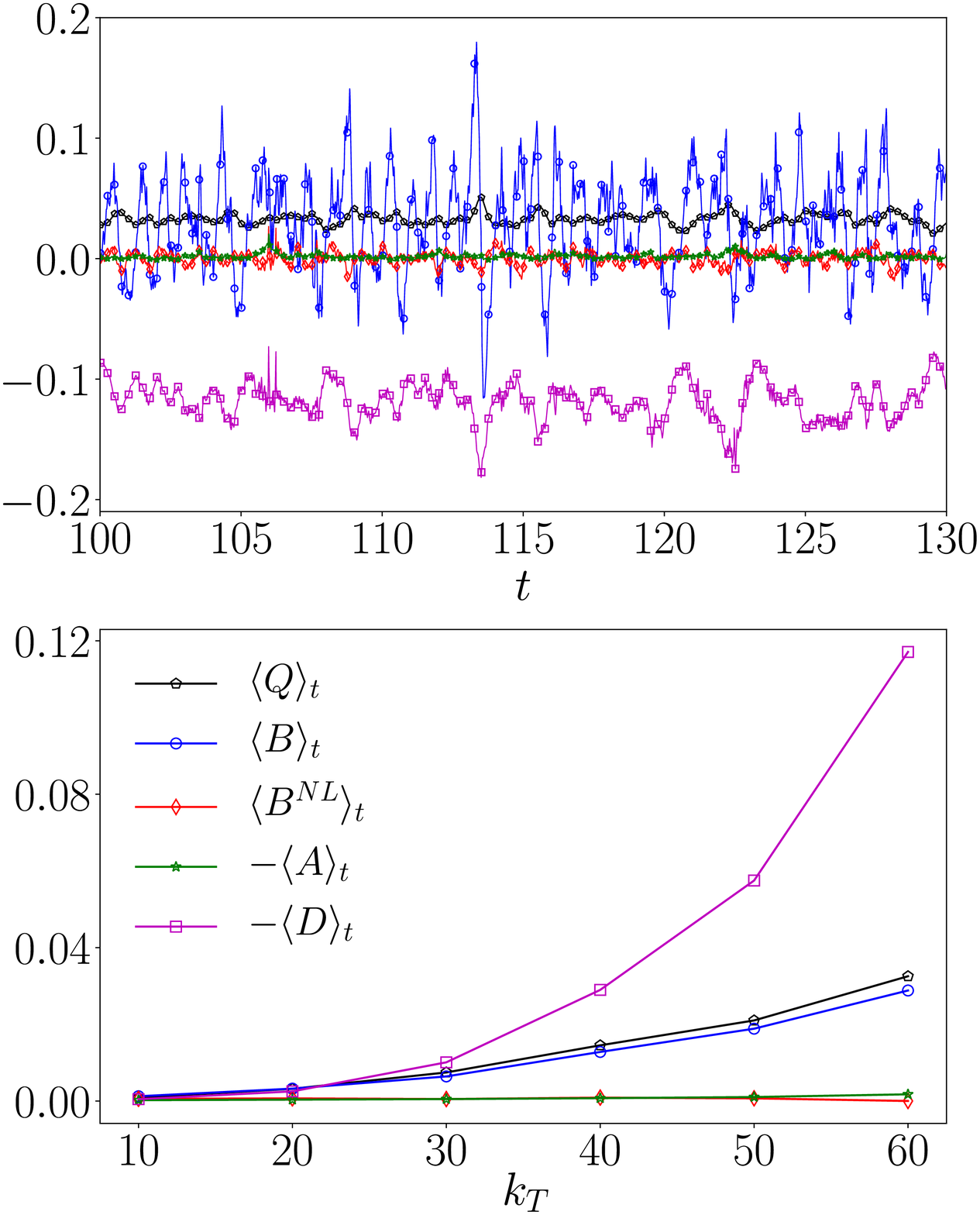}
\put(-255,320){(a)}
\put(-255,160){(b)}
 \caption{\Revision{}{(a) Time series of total enstrophy $\sum_k \widehat{Q}_k$, baroclinic term $\sum_k \widehat{B}_k$, nonlinear baroclinic term $\sum_k\widehat{B}^{NL}_k$, dissipation $\sum_k \widehat{D}_k$, and the advection $\sum_k \widehat{A}_k$ term for simulation 6d$^r$. The baroclinic production oscillates about a positive mean value (averaged in time). (b) Time averaged total enstrophy $\langle Q \rangle_t$, baroclinic term $\langle B \rangle_t$, nonlinear baroclinic term $\langle B^{NL}\rangle_t$ negative of dissipation $\langle D\rangle_t$, and advection $\langle A\rangle_t$ terms. Magnitude of $\langle D \rangle_t$ is higher than $\langle B \rangle_t$. Difference between the two is accounted for by the energy injection due to energy rescaling $\widehat{F}^r_k$. Legends in both (a) and (b) are same.}}
 \label{fig: EnstrophyBudgetTimeSeries}
\end{figure}
\begin{figure}[!t]
\centering
 \includegraphics[width=0.5\textwidth]{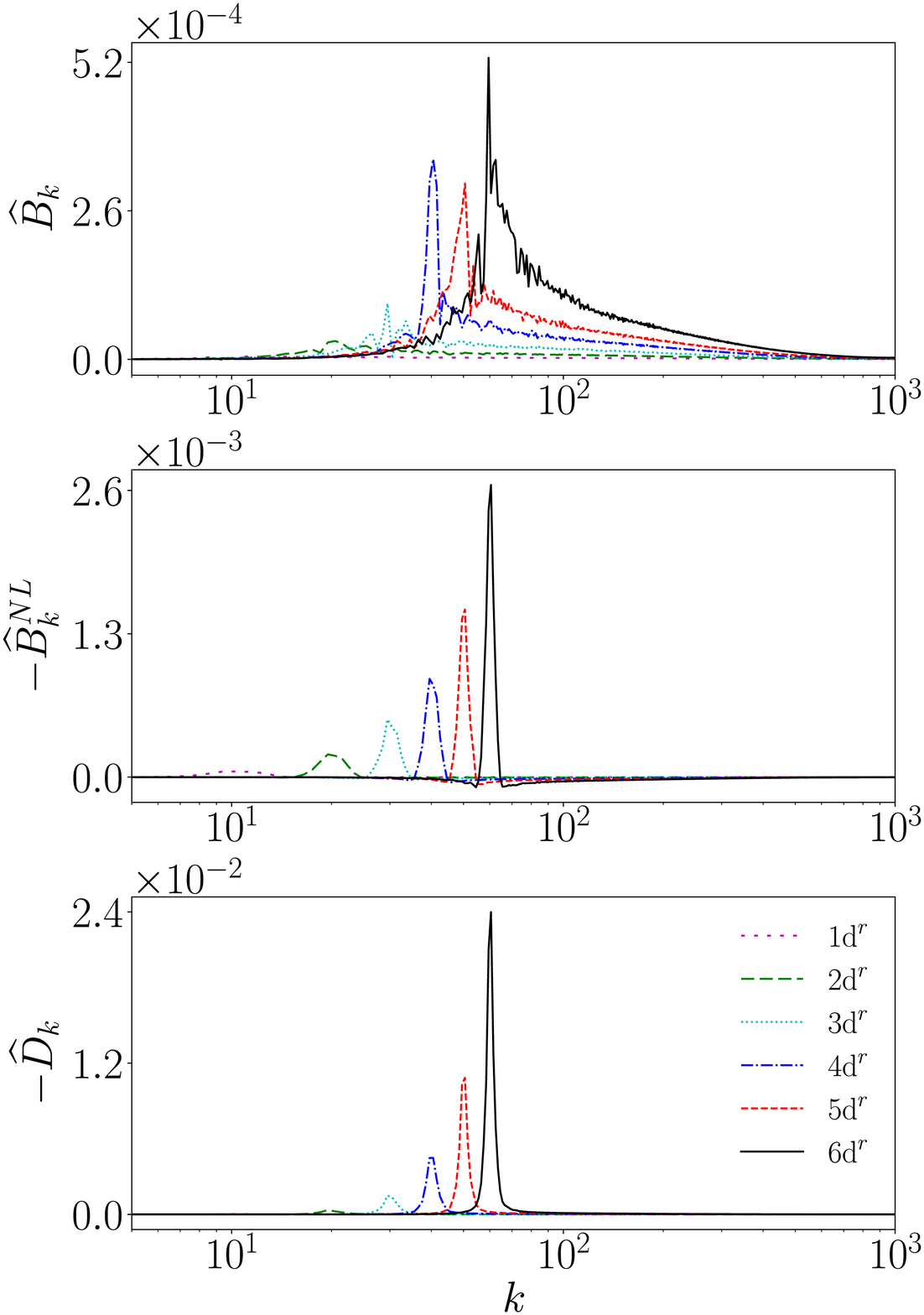}
\put(-250,350){(a)}
 \put(-250,235){(b)}
\put(-250,120){(c)}
 \caption{\Revision{}{(a) Spectra of baroclinic term $\widehat{B}_k$, negative of nonlinear baroclinic term $\widehat{B}^{NL}_k$, and the negative of (c) dissipation term $-\widehat{D}_k$ in enstrophy Eq.~\eqref{eq: EnstrophySpectra}. }}
 \label{fig: EnstrophyBudgetSpectra}
\end{figure}

First term on the right of Eq.~\eqref{eq: EnstrophySpectra}, $\widehat{B}_k$, denotes the baroclinic generation of enstrophy and captures the effect of the baroclinic torque on enstrophy. The next three terms (combined in $\widehat{D}_k$) appear due to the viscous forces in the momentum equation. \Revision{We note that not}{Not} all three terms corresponding to \Revision{}{the} viscous forces necessarily destroy enstrophy. The first term of these three terms which is proportional to \Revision{}{the} Laplacian of vorticity is always non-positive and has primary contribution in viscous dissipation of enstrophy. The second term and the third term of these three terms can be positive as well as negative. The next term ($\widehat{A}_k$) arises due to convection term in the momentum equation and has very small contribution in changing enstrophy of the system. Lastly, $\widehat{F}^r_k$ is due to density-weighted energy rescaling. As discussed in Section~\ref{sec: Vorticity}, a simplified form of energy rescaling as a right hand side term in governing equations is difficult so we do not simplify $\widehat{F}^r_k$. 

Figure~\ref{fig: EnstrophySpectra} shows the spectra of enstrophy for simulations with energy rescaling. As noted for vortical energy spectra $\widehat{E}^f_{vk}$ in Fig.~\ref{fig: EfvVSEvRescaled}, the vorticity peaks at the same wavenumber at which background thermal gradients exist ($k_T$). Consequently, the enstrophy spectra also peak at $k_T$. Such localized (in the spectral space) enstrophy is generated due to both the baroclinic production term $\widehat{B}_k$ as well as the energy rescaling term $\widehat{F}^r_k$. Figure~\ref{fig: EnstrophyBudgetTimeSeries}(a) shows the time series of right hand side terms of Eq.~\eqref{eq: EnstrophySpectra} and total enstrophy for simulation 6d$^r$. The baroclinic term oscillates about a positive mean value while the advection terms have negligible contribution to the change of enstrophy in time. Figure~\ref{fig: EnstrophyBudgetTimeSeries}(b) shows the time averaged values of the enstrophy, the full baroclinic production term and the nonlinear baroclinic production term, negative of the dissipation, and the advection terms for simulations 1d$^r$-6d$^r$ ($\langle\cdot\rangle_t$ denoting time average). As the wavenumber $k_T$ increases, \Revision{we note that}{} the enstrophy in the system increases. Accordingly, the baroclinic generation and the dissipation also increase. \Revision{}{Enstrophy $\widehat{Q}_k$ is statistically stationary since the vortical energy $E_v$ is statistically stationary as shown in Fig.~\ref{fig: TimeSeriesEwEv}(b). Consequently, $\widehat{F}^r_k$ accounts for the difference between the two since the L.H.S of Eq.~\eqref{eq: EnstrophySpectra} vanishes.}
Density gradients in curved shock waves need not be parallel to the pressure gradients due to thermoviscous generation of entropy within the shock waves. Furthermore, due to imposed thermal inhomogeneity, shock waves get reflected and refracted continuously in the domain. A combination of these two effects can be seen by decomposing the baroclinic term $\widehat{B}_k$ as, 
\begin{equation}
 \widehat{B}_k = \mathrm{Re}\left[\widehat{\bfOmega}^*_k\cdot\widehat{\left(\frac{\nabla f \times \nabla p}{\rho^2}\right)}_k\right] + \underbrace{\mathrm{Re}\left[\widehat{\bfOmega}^*_k\cdot\widehat{\left(\frac{\nabla (\rho - f) \times \nabla p}{\rho^2}\right)}_k\right]}_{\widehat{B}^{NL}_k},
 \label{eq: baroclinicNonlinear}
\end{equation}
where $\widehat{B}^{NL}_k$ captures the nonlinear baroclinic interactions between pressure gradients and density gradients. Figure~\ref{fig: EnstrophyBudgetTimeSeries}(a) shows the time series of the nonlinear baroclinic term for simulation 6d$^r$ and Fig.~\ref{fig: EnstrophyBudgetTimeSeries}(b), time average of the nonlinear baroclinic interaction term is shown. The overall contribution of the nonlinear baroclinic term to enstrophy is very small. However, as shown in Fig.~\ref{fig: EnstrophyBudgetSpectra}(b), the nonlinear baroclinic interaction is local in the wavenumber space and negative in value, indicating its contribution in destruction of enstrophy. Additionally, the nonlinear baroclinic term peaks near $k_T$, indicating that any nonlinear baroclinic interaction is amplified due to the background thermal gradients. Similarly, the dissipation term $\widehat{D}_k$ is also highly negative and localized near $k_T$. On the other hand, the total baroclinic term $\widehat{B}_k$ has a nonlocal contribution indicating the participation of a wide range of length scales in generation of enstrophy. As shown in Fig.~\ref{fig: EnstrophyBudgetTimeSeries}(a) and~\ref{fig: EnstrophyBudgetTimeSeries}(b), the overall contribution of $\widehat{B}_k$ is positive and large. Furthermore, the difference between $\widehat{B}_k$ and $\widehat{D}_k$ in Fig.~\ref{fig: EnstrophyBudgetTimeSeries}(b) represents $\widehat{F}^r_k$ in Eq.~\eqref{eq: EnstrophySpectra}. In Appendix~\ref{app: diss}, we discuss the relative magnitudes of dissipation terms in Eq.~\eqref{eq: enstrophy_dissipation}.

\section{Conclusions}
\label{sec: conclusions}
We have studied forced two-dimensional shock waves in a periodic domain with an inhomogeneous background temperature field using numerical simulations. We showed that random forcing generates an isotropic field of shock waves and the density-weighted wave energy spectra scale with wavenumber as $k^{-2}$ as in Eq.~\eqref{eq: energySpectraScaling}. The flux of energy spectra does not remain constant in the cascade range of wavenumbers if physical diffusion is considered indicating the spectral energy transfer to be of variable flux type~\cite{verma2021variable}. We have also shown that  dissipation exists in all length scales, unlike hydrodynamic turbulence, where dissipation is localized in smaller length scales. Such energy spectra cascade characteristics remain similar when the field of shock waves propagates in an inhomogeneous medium with randomized thermal gradients. 

While the interaction of shock waves with background thermal gradients results in a baroclinic torque generating vorticity, the type of energy injection in shock waves governs the evolution of this vorticity field in the domain. We showed that for stochastically forced shock waves, no coherent vorticity is generated, and only a broadband vorticity (distributed over all length scales) exists. For stochastic forcing, the pressure gradient field in the domain is also a stochastic process. Since the direction of the pressure gradient is also stochastic, time-averaged baroclinic torque vanishes since the leading order term is linearly dependent on the pressure gradient. Consequently, only broadband vorticity is observed due to changing direction of pressure gradients. The broadband vorticity is high in the forced length scales and decays for smaller length scales, further highlighting that it is generated due to the stochastic forcing. To investigate the generation of coherent vorticity, we have introduced a new forcing based on wave energy rescaling. In this forcing, we rescale the energy in the density-weighted wave modes at each time step. The cascade is thus driven by waves that do not change phase and propagation direction continuously. Consequently, coherent vorticity is generated in the domain at the same length scale as that of the thermal gradients. We have also shown that density-weighted variables introduced by~\citet{miura1995acoustic} are not suitable for studying rotational and dilatational flow fields separately in compressible flows. \RevisionTwo{}{Though in the current study we have investigated the baroclinic interaction of forced weak shock waves, we expect the baroclinic vorticity production to increase for stronger shock waves due to higher pressure gradients. However, such simulations are beyond the scope of the current study. }\Revision{}{For all the simulations we used constant viscosity and thermal conductivity, however, as shown in Appendix~\ref{app: hyper}, temperature dependent viscosity and conductivity have negligible affects on the energy spectra of shock waves. Consequently, the baroclinic interaction is not expected to be affected by the temperature dependent viscosity and conductivity values in the weak-shock regime.}

In practice, the generation of random weak shock waves may be due to intense \Revision{}{unsteady} combustion events or thermoacoustic instabilities. Spatially distributed \Revision{}{unsteady combustion events are better represented by stochastic forcing due to randomly generated pressure waves~\cite{dowling2015combustion}}, while instability-driven shock waves are better represented by forcing based on energy rescaling since only particular modes exhibit such instabilities~\cite{gupta2017spectral} \Revision{}{depending on the geometry of the combustion chamber}. \Revision{}{Besides the practical relevance in understanding unsteady shock induced mixing, we also highlighted that the method of injection of energy in the so-called \emph{wave-turbulent}~\cite{nazarenko2011wave} systems in which nonlinear waves interact with each other is a richer problem compared to the hydrodynamic turbulent systems. In hydrodynamic turbulence, there is no wave velocity of propagation of eddies due to which phasing is never considered in hydrodynamic turbulence forcing~\cite{rosales2005linear}. However, phasing or phase difference between pressure and velocity in both space and time is an important parameter governing the acoustic wave turbulence and interaction with background medium.} Accordingly, future research directions include a detailed investigation of random shock wave forcing and the possibility of sustaining hydrodynamic turbulence via forced weak shock waves in three dimensions.

\begin{acknowledgments}
We acknowledge the financial support received from Science and Engineering Research Board (SERB), Government of India under Grant No. SRG/2022/000728. We also thank IIT Delhi HPC facility for computational resources.
\end{acknowledgments}
\appendix
\section{Effects of hyperdiffusion and variable viscosity}
\label{app: hyper}
\begin{figure}[!b]
        \centering
        \includegraphics[width=0.5\textwidth]{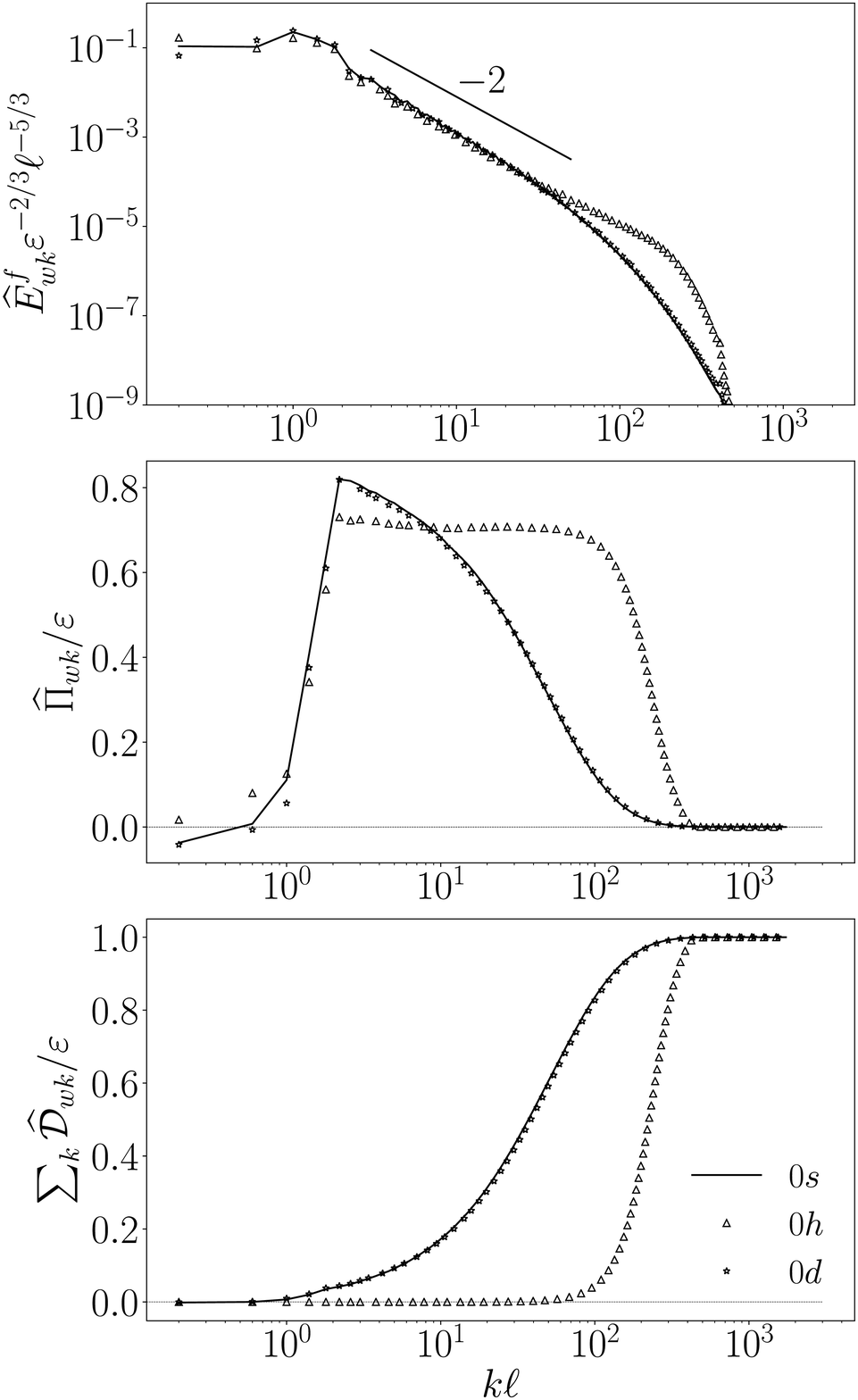}
        \put(-250,415){(a)}
        \put(-250,280){(b)}
        \put(-250,140){(c)}
        \caption{\Revision{}{(a) Dimensionless scaled density-weighted wave energy spectra $\widehat{E}^f_{wk}$, (b) spectral flux of wave energy $\widehat{\Pi}_{wk}$, and (c) cumulative dissipation $\widehat{\mathcal{D}}_{wk}$ for a baseline case (0d in Table~\ref{tab: baselineCases}), compared against a simulation with viscosity varying with the Sutherland law in Eq.~\eqref{eq: appSutherland} (0s) and a simulation with hyperdiffusion (0h). For simulation 0s, all parameters are kept same as 0d simulation except for viscosity and thermal conductivity. Prandtl number $\mathrm{Pr}$ is fixed at 0.72. For simulation 0h, hyperdiffusion is used with $r=2$ and $\nu = 5.44\times 10^{-10}$ (c.f. Eqs.~\eqref{eq: appHyperMomentum} and \eqref{eq: appHyperPressure}).}}
        \label{fig: baselineSpectraHyper}
    \end{figure}
Throughout our simulations, we consider the dimensionless viscosity constant ($\mu=1$). However, for high thermal perturbations, changes in viscosity due to inhomogeneous temperature may be significant. To confirm that such thermal dependence of viscosity has an insignificant impact on the spectral wave energy cascade in our simulations, we also perform a simulation with dimensionless viscosity governed by the Sutherland's law 
\begin{equation}
 \mu = T^{3/2}\left(\frac{T_m + S_{ref}}{T T_m + S_{ref}}\right),
 \label{eq: appSutherland}
\end{equation}
where $T_m = 273.15~\mathrm{K}$ and is the reference temeperature scale (see Eq.~\eqref{eq: referenceScales}), $S_{ref} = 110.4~\mathrm{K}$, and $T$ is the dimensionless temperature. \RevisionTwo{We also note that to}{To} obtain a wide range of length scales in a given resolution, hyperdiffusion is used in wave-turbulence studies~\cite{augier2019shallow, farge1989wave}. Such hyperdiffusion modifies all the viscous stress tensor and thermal conduction terms in the momentum and pressure governing Eqs.~\eqref{eq: momentum_equation} and~\eqref{eq: pressure_equation} as, 
\begin{align}
 \frac{\partial \bfu}{\partial t} + \bfu\cdot\nabla\bfu = -\frac{\nabla p}{\rho} + \nu\nabla^{2r}\bfu,\label{eq: appHyperMomentum}\\
 \frac{\partial p}{\partial t} + \bfu\cdot\nabla p + \gamma p \nabla\cdot\bfu = \nu\nabla^{2r}p \label{eq: appHyperPressure},
\end{align}
where $r$ and $\nu$ are parameters of hyperdiffusivity. To assert the validity of constant dynamic viscosity, we compare the energy spectra of simulation 0d in Table~\ref{tab: baselineCases}, with two simulations, one with viscosity varying with local temperature as per Eq.~\eqref{eq: appSutherland} (simulation 0s) and one with hyperdiffusion with $r=2$ and $\nu=5.44\times 10^{-10}$ (simulation 0h) in Fig.~\ref{fig: baselineSpectraHyper}.

\Revision{We note that}{As shown in Fig.~\ref{fig: baselineSpectraHyper}} temperature dependent viscosity does not affect the energy spectra significantly. For simulation with hyperdiffusion (0h), the energy spectra extend with a $-2$ slope for larger range of wavenumbers. However, the flux of spectral energy $\widehat{\Pi}$ and cumulative spectral dissipation $\widehat{\mathcal{D}}_k$ are modified significantly. \Revision{We note that the}{The} flux is approximately constant in the \Revision{spectra transfer range}{spectral transfer range} and the dissipation is accumulated in higher wavenumbers. However, due to physical viscous terms, the spectra flux decays smoothly and dissipation exists in all wavenumbers. \Revision{We further note that,}{Furthermore,} even in a hypothetical simulations with infinite resolution, since the energy spectra will decay as $k^{-2}$, the dissipation will exist in all scales (since $\widehat{\mathcal{D}}_k\sim k^2\widehat{E}_k$). Hence, replacing physical dissipation terms with a hyperdiffusion term tends to modify the fundamental nature of spectral energy transfer. 
Since the variation of viscosity has negligible effect on energy spectra and hyperdiffusion modifies the fundamental nature of spectral energy transfer in nonlinear acoustic wave turbulence, we choose dimensionless viscosity $\mu=1$ in all our simulations.
\section{Enstrophy dissipation}
\label{app: diss}
\begin{figure}[!t]
        \centering
        \includegraphics[width=0.5\textwidth]{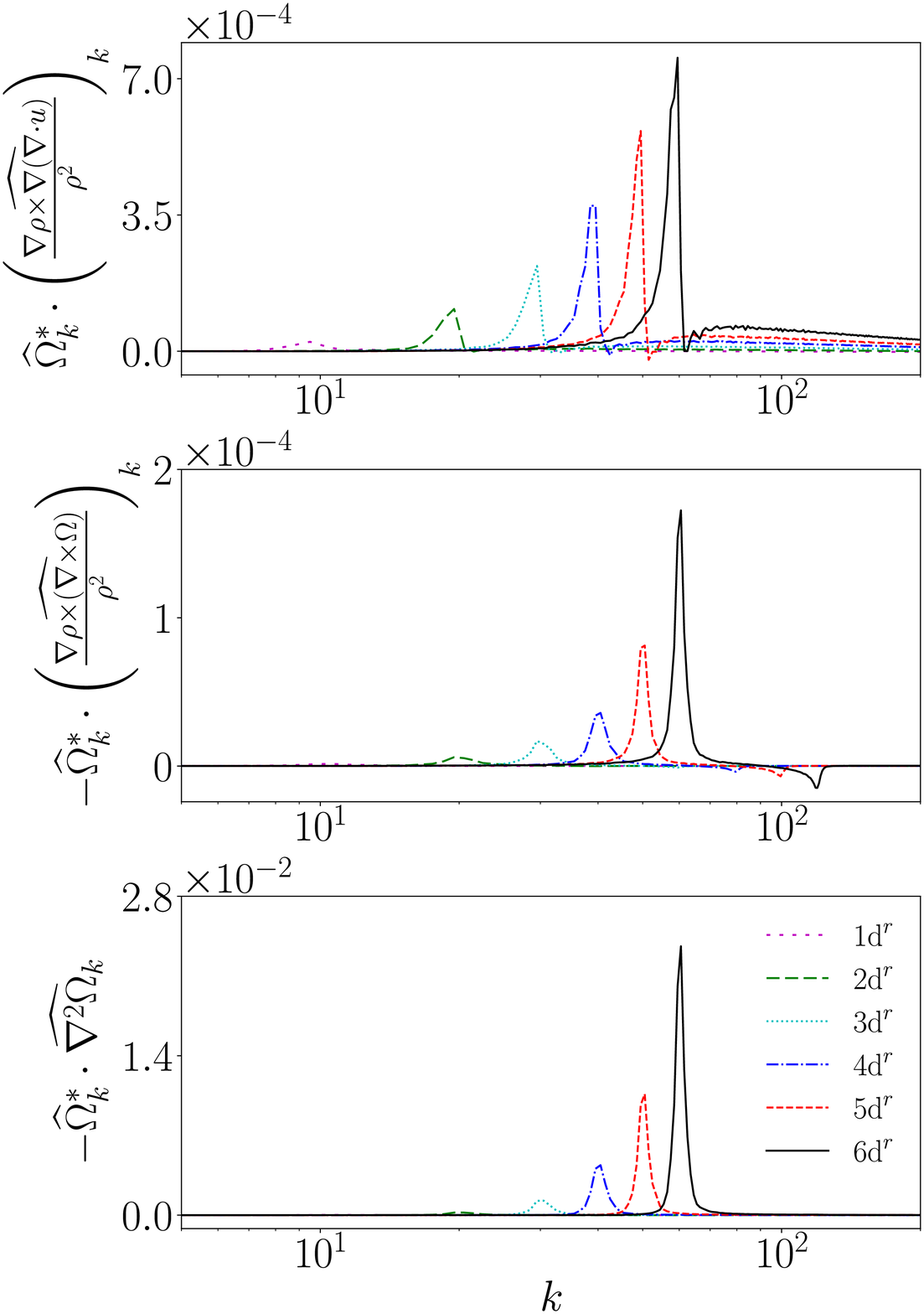}
        \put(-250,360){(a)}
        \put(-250,240){(b)}
        \put(-250,120){(c)}
        \caption{\Revision{}{(a) Spectra of dissipation due to density variation interacting with velocity divergence, (b) curl of vorticity, and (c) Laplacian of vorticity as derived in Eq.~\eqref{eq: enstrophy_dissipation}.}}
        \label{fig: enstrophy_dissipation_spectra}
    \end{figure}
Three terms contribute to enstrophy dissipation as derived in Eq.~\eqref{eq: enstrophy_dissipation}. In Fig.~\ref{fig: enstrophy_dissipation_spectra} we show the spectra of all three terms contributing to the dissipation of enstrophy in simulations 1d$^r$-6d$^r$ (see Table~\ref{tab: parameterSpace}). Similar to vorticity and enstrophy, all the dissipation terms are localized around $k_T$ signifying coherent vorticity in the domain. Figure~\ref{fig: enstrophy_dissipation_spectra}(a) shows spectra of dissipation due to misalignment between density gradient and bulk viscous forces (quantified by $\nabla\left(\nabla\cdot\bfu\right)$. Figure~\ref{fig: enstrophy_dissipation_spectra}b shows the spectra of dissipation due to misalignment between density gradient and a combination of bulk and shear viscous forces (since $\nabla\times\Omega = \nabla\left(\nabla\cdot\bfu\right) - \nabla^2\bfu$). Together, the terms shown in Figs.~\ref{fig: enstrophy_dissipation_spectra}(a) and~\ref{fig: enstrophy_dissipation_spectra}(b) maybe considered as viscous counterparts of baroclinic torque in which misalignment of density gradients with viscous forces contributes to dissipation. \RevisionTwo{It is important to note that}{We highlight that} these two terms are not necessarily dissipative in nature~\cite{kida1990enstrophy}. Figure~\ref{fig: enstrophy_dissipation_spectra}(c) shows the spectra of the Laplacian dissipation of vorticity and is necessarily negative. \Revision{We note that the}{The} Laplacian dissipation is highest in rescaled simulations denoting highly localized (in spectral space) vorticity. Since the vorticity and production of enstrophy are highly sensitive to the method of forcing random shock waves (stochastic forcing compared with rescaled forcing in this work), we expect the enstrophy dissipation to also change with the forcing methodology. 

\bibliography{references}

\end{document}

%% file: defs.tex
\newcommand{\boldface}[1]{\boldsymbol{#1}}  
\newcommand{\bfa}{\boldface{a}}

\newcommand{\bfk}{\boldface{k}}

\newcommand{\bfp}{\boldface{p}}

\newcommand{\bfu}{\boldface{u}}

\newcommand{\bfw}{\boldface{w}}

\newcommand{\bfD}{\boldface{D}}

\newcommand{\bfF}{\boldface{F}}

\newcommand{\bfI}{\boldface{I}}

\newcommand{\bfS}{\boldface{S}}

\newcommand{\bfphi}{\boldsymbol{\phi}}

\newcommand{\bfOmega}{\boldsymbol{\Omega}}

\newcommand{\T}{^{\mathrm{T}}} 

\newlength{\boxwidth}
\def\btheorem{\begin{theorem}}
\def\etheorem{\end{theorem}}
\def\blemma{\begin{lemma}}
\def\elemma{\end{lemma}}
\def\bproposition{\begin{proposition}}
\def\eproposition{\end{proposition}}
\def\bcorollary{\begin{corollary}}
\def\ecorollary{\end{corollary}}
\def\bdefinition{\begin{definition}}
\def\edefinition{\end{definition}}
\def\bexample{\begin{example}}
\def\eexample{\end{example}}
\def\bremark{\begin{remark}}
\def\eremark{\end{remark}}

\newcommand{\be}{\begin{equation*}}
\newcommand{\ee}{\end{equation*}}

\newcommand{\beq}{\begin{eqnarray*}}
\newcommand{\eeq}{\end{eqnarray*}}
\newcommand{\bem}{\begin{multline}}
\newcommand{\eem}{\end{multline}}
\newcommand{\ba}{\begin{align*}}
\newcommand{\ea}{\end{align*}}

\renewcommand{\figurename}{Figure}
\renewcommand{\tablename}{Table}

%% file: JossyGupta_2023.bbl
\providecommand{\noopsort}[1]{}\providecommand{\singleletter}[1]{#1}%
\begin{thebibliography}{47}%
\makeatletter
\providecommand \@ifxundefined [1]{%
 \@ifx{#1\undefined}
}%
\providecommand \@ifnum [1]{%
 \ifnum #1\expandafter \@firstoftwo
 \else \expandafter \@secondoftwo
 \fi
}%
\providecommand \@ifx [1]{%
 \ifx #1\expandafter \@firstoftwo
 \else \expandafter \@secondoftwo
 \fi
}%
\providecommand \natexlab [1]{#1}%
\providecommand \enquote  [1]{``#1''}%
\providecommand \bibnamefont  [1]{#1}%
\providecommand \bibfnamefont [1]{#1}%
\providecommand \citenamefont [1]{#1}%
\providecommand \href@noop [0]{\@secondoftwo}%
\providecommand \href [0]{\begingroup \@sanitize@url \@href}%
\providecommand \@href[1]{\@@startlink{#1}\@@href}%
\providecommand \@@href[1]{\endgroup#1\@@endlink}%
\providecommand \@sanitize@url [0]{\catcode `\\12\catcode `\$12\catcode
  `\&12\catcode `\#12\catcode `\^12\catcode `\_12\catcode `\%12\relax}%
\providecommand \@@startlink[1]{}%
\providecommand \@@endlink[0]{}%
\providecommand \url  [0]{\begingroup\@sanitize@url \@url }%
\providecommand \@url [1]{\endgroup\@href {#1}{\urlprefix }}%
\providecommand \urlprefix  [0]{URL }%
\providecommand \Eprint [0]{\href }%
\providecommand \doibase [0]{http://dx.doi.org/}%
\providecommand \selectlanguage [0]{\@gobble}%
\providecommand \bibinfo  [0]{\@secondoftwo}%
\providecommand \bibfield  [0]{\@secondoftwo}%
\providecommand \translation [1]{[#1]}%
\providecommand \BibitemOpen [0]{}%
\providecommand \bibitemStop [0]{}%
\providecommand \bibitemNoStop [0]{.\EOS\space}%
\providecommand \EOS [0]{\spacefactor3000\relax}%
\providecommand \BibitemShut  [1]{\csname bibitem#1\endcsname}%
\let\auto@bib@innerbib\@empty
\bibitem [{\citenamefont {Goldstein}\ and\ \citenamefont
  {Wundrow}(1990)}]{goldstein1990spatial}%
  \BibitemOpen
  \bibfield  {author} {\bibinfo {author} {\bibfnamefont {M.}~\bibnamefont
  {Goldstein}}\ and\ \bibinfo {author} {\bibfnamefont {D.}~\bibnamefont
  {Wundrow}},\ }\bibfield  {title} {\enquote {\bibinfo {title} {Spatial
  evolution of nonlinear acoustic mode instabilities on hypersonic boundary
  layers},}\ }\href@noop {} {\bibfield  {journal} {\bibinfo  {journal} {Journal
  of Fluid Mechanics}\ }\textbf {\bibinfo {volume} {219}},\ \bibinfo {pages}
  {585--607} (\bibinfo {year} {1990})}\BibitemShut {NoStop}%
\bibitem [{\citenamefont {Liepmann}\ and\ \citenamefont
  {Roshko}(2001)}]{liepmann2001elements}%
  \BibitemOpen
  \bibfield  {author} {\bibinfo {author} {\bibfnamefont {H.~W.}\ \bibnamefont
  {Liepmann}}\ and\ \bibinfo {author} {\bibfnamefont {A.}~\bibnamefont
  {Roshko}},\ }\href@noop {} {\emph {\bibinfo {title} {Elements of
  gasdynamics}}}\ (\bibinfo  {publisher} {Courier Corporation},\ \bibinfo
  {year} {2001})\BibitemShut {NoStop}%
\bibitem [{\citenamefont {Baars}\ and\ \citenamefont
  {Tinney}(2013)}]{baars2013transient}%
  \BibitemOpen
  \bibfield  {author} {\bibinfo {author} {\bibfnamefont {W.}~\bibnamefont
  {Baars}}\ and\ \bibinfo {author} {\bibfnamefont {C.}~\bibnamefont {Tinney}},\
  }\bibfield  {title} {\enquote {\bibinfo {title} {Transient wall pressures in
  an overexpanded and large area ratio nozzle},}\ }\href@noop {} {\bibfield
  {journal} {\bibinfo  {journal} {Experiments in fluids}\ }\textbf {\bibinfo
  {volume} {54}},\ \bibinfo {pages} {1--17} (\bibinfo {year}
  {2013})}\BibitemShut {NoStop}%
\bibitem [{\citenamefont {Bonciolini}, \citenamefont {Boujo},\ and\
  \citenamefont {Noiray}(2017)}]{bonciolini2017output}%
  \BibitemOpen
  \bibfield  {author} {\bibinfo {author} {\bibfnamefont {G.}~\bibnamefont
  {Bonciolini}}, \bibinfo {author} {\bibfnamefont {E.}~\bibnamefont {Boujo}}, \
  and\ \bibinfo {author} {\bibfnamefont {N.}~\bibnamefont {Noiray}},\
  }\bibfield  {title} {\enquote {\bibinfo {title} {Output-only parameter
  identification of a colored-noise-driven van-der-pol oscillator:
  thermoacoustic instabilities as an example},}\ }\href@noop {} {\bibfield
  {journal} {\bibinfo  {journal} {Physical Review E}\ }\textbf {\bibinfo
  {volume} {95}},\ \bibinfo {pages} {062217} (\bibinfo {year}
  {2017})}\BibitemShut {NoStop}%
\bibitem [{\citenamefont {Gupta}, \citenamefont {Lodato},\ and\ \citenamefont
  {Scalo}(2017)}]{gupta2017spectral}%
  \BibitemOpen
  \bibfield  {author} {\bibinfo {author} {\bibfnamefont {P.}~\bibnamefont
  {Gupta}}, \bibinfo {author} {\bibfnamefont {G.}~\bibnamefont {Lodato}}, \
  and\ \bibinfo {author} {\bibfnamefont {C.}~\bibnamefont {Scalo}},\ }\bibfield
   {title} {\enquote {\bibinfo {title} {Spectral energy cascade in
  thermoacoustic shock waves},}\ }\href@noop {} {\bibfield  {journal} {\bibinfo
   {journal} {Journal of Fluid Mechanics}\ }\textbf {\bibinfo {volume} {831}},\
  \bibinfo {pages} {358--393} (\bibinfo {year} {2017})}\BibitemShut {NoStop}%
\bibitem [{\citenamefont {Breazeale}\ and\ \citenamefont
  {Thompson}(1963)}]{breazeale1963finite}%
  \BibitemOpen
  \bibfield  {author} {\bibinfo {author} {\bibfnamefont {M.}~\bibnamefont
  {Breazeale}}\ and\ \bibinfo {author} {\bibfnamefont {D.}~\bibnamefont
  {Thompson}},\ }\bibfield  {title} {\enquote {\bibinfo {title}
  {Finite-amplitude ultrasonic waves in aluminum},}\ }\href@noop {} {\bibfield
  {journal} {\bibinfo  {journal} {Applied Physics Letters}\ }\textbf {\bibinfo
  {volume} {3}},\ \bibinfo {pages} {77--78} (\bibinfo {year}
  {1963})}\BibitemShut {NoStop}%
\bibitem [{\citenamefont {Lissenden}(2021)}]{lissenden2021nonlinear}%
  \BibitemOpen
  \bibfield  {author} {\bibinfo {author} {\bibfnamefont {C.~J.}\ \bibnamefont
  {Lissenden}},\ }\bibfield  {title} {\enquote {\bibinfo {title} {Nonlinear
  ultrasonic guided waves—principles for nondestructive evaluation},}\
  }\href@noop {} {\bibfield  {journal} {\bibinfo  {journal} {Journal of Applied
  Physics}\ }\textbf {\bibinfo {volume} {129}},\ \bibinfo {pages} {021101}
  (\bibinfo {year} {2021})}\BibitemShut {NoStop}%
\bibitem [{\citenamefont {Sagdeev}(1979)}]{sagdeev19791976}%
  \BibitemOpen
  \bibfield  {author} {\bibinfo {author} {\bibfnamefont {R.~Z.}\ \bibnamefont
  {Sagdeev}},\ }\bibfield  {title} {\enquote {\bibinfo {title} {The 1976
  oppenheimer lectures: Critical problems in plasma astrophysics. i. turbulence
  and nonlinear waves},}\ }\href@noop {} {\bibfield  {journal} {\bibinfo
  {journal} {Reviews of Modern Physics}\ }\textbf {\bibinfo {volume} {51}},\
  \bibinfo {pages} {1} (\bibinfo {year} {1979})}\BibitemShut {NoStop}%
\bibitem [{\citenamefont {Brouillette}(2002)}]{brouillette2002richtmyer}%
  \BibitemOpen
  \bibfield  {author} {\bibinfo {author} {\bibfnamefont {M.}~\bibnamefont
  {Brouillette}},\ }\bibfield  {title} {\enquote {\bibinfo {title} {The
  richtmyer-meshkov instability},}\ }\href@noop {} {\bibfield  {journal}
  {\bibinfo  {journal} {Annual Review of Fluid Mechanics}\ }\textbf {\bibinfo
  {volume} {34}},\ \bibinfo {pages} {445--468} (\bibinfo {year}
  {2002})}\BibitemShut {NoStop}%
\bibitem [{\citenamefont {Lukes}\ \emph {et~al.}(2016)\citenamefont {Lukes},
  \citenamefont {Fern{\'a}ndez}, \citenamefont {Guti{\'e}rrez-Aceves},
  \citenamefont {Fern{\'a}ndez}, \citenamefont {Alvarez}, \citenamefont
  {Sunka},\ and\ \citenamefont {Loske}}]{lukes2016tandem}%
  \BibitemOpen
  \bibfield  {author} {\bibinfo {author} {\bibfnamefont {P.}~\bibnamefont
  {Lukes}}, \bibinfo {author} {\bibfnamefont {F.}~\bibnamefont
  {Fern{\'a}ndez}}, \bibinfo {author} {\bibfnamefont {J.}~\bibnamefont
  {Guti{\'e}rrez-Aceves}}, \bibinfo {author} {\bibfnamefont {E.}~\bibnamefont
  {Fern{\'a}ndez}}, \bibinfo {author} {\bibfnamefont {U.}~\bibnamefont
  {Alvarez}}, \bibinfo {author} {\bibfnamefont {P.}~\bibnamefont {Sunka}}, \
  and\ \bibinfo {author} {\bibfnamefont {A.}~\bibnamefont {Loske}},\ }\bibfield
   {title} {\enquote {\bibinfo {title} {Tandem shock waves in medicine and
  biology: a review of potential applications and successes},}\ }\href@noop {}
  {\bibfield  {journal} {\bibinfo  {journal} {Shock waves}\ }\textbf {\bibinfo
  {volume} {26}},\ \bibinfo {pages} {1--23} (\bibinfo {year}
  {2016})}\BibitemShut {NoStop}%
\bibitem [{\citenamefont {Lighthill}(1978)}]{lighthill1978acoustic}%
  \BibitemOpen
  \bibfield  {author} {\bibinfo {author} {\bibfnamefont {J.}~\bibnamefont
  {Lighthill}},\ }\bibfield  {title} {\enquote {\bibinfo {title} {Acoustic
  streaming},}\ }\href@noop {} {\bibfield  {journal} {\bibinfo  {journal}
  {Journal of sound and vibration}\ }\textbf {\bibinfo {volume} {61}},\
  \bibinfo {pages} {391--418} (\bibinfo {year} {1978})}\BibitemShut {NoStop}%
\bibitem [{\citenamefont {Shimizu}\ and\ \citenamefont
  {Sugimoto}(2010)}]{shimizu2010numerical}%
  \BibitemOpen
  \bibfield  {author} {\bibinfo {author} {\bibfnamefont {D.}~\bibnamefont
  {Shimizu}}\ and\ \bibinfo {author} {\bibfnamefont {N.}~\bibnamefont
  {Sugimoto}},\ }\bibfield  {title} {\enquote {\bibinfo {title} {Numerical
  study of thermoacoustic taconis oscillations},}\ }\href@noop {} {\bibfield
  {journal} {\bibinfo  {journal} {Journal of Applied Physics}\ }\textbf
  {\bibinfo {volume} {107}},\ \bibinfo {pages} {034910} (\bibinfo {year}
  {2010})}\BibitemShut {NoStop}%
\bibitem [{\citenamefont {Whitham}(2011)}]{whitham2011linear}%
  \BibitemOpen
  \bibfield  {author} {\bibinfo {author} {\bibfnamefont {G.~B.}\ \bibnamefont
  {Whitham}},\ }\href@noop {} {\emph {\bibinfo {title} {Linear and nonlinear
  waves}}}\ (\bibinfo  {publisher} {John Wiley \& Sons},\ \bibinfo {year}
  {2011})\BibitemShut {NoStop}%
\bibitem [{\citenamefont {Gupta}\ and\ \citenamefont
  {Scalo}(2018)}]{gupta2018spectral}%
  \BibitemOpen
  \bibfield  {author} {\bibinfo {author} {\bibfnamefont {P.}~\bibnamefont
  {Gupta}}\ and\ \bibinfo {author} {\bibfnamefont {C.}~\bibnamefont {Scalo}},\
  }\bibfield  {title} {\enquote {\bibinfo {title} {Spectral energy cascade and
  decay in nonlinear acoustic waves},}\ }\href@noop {} {\bibfield  {journal}
  {\bibinfo  {journal} {Physical Review E}\ }\textbf {\bibinfo {volume} {98}},\
  \bibinfo {pages} {033117} (\bibinfo {year} {2018})}\BibitemShut {NoStop}%
\bibitem [{\citenamefont {Thirani}, \citenamefont {Gupta},\ and\ \citenamefont
  {Scalo}(2020)}]{thirani2020knudsen}%
  \BibitemOpen
  \bibfield  {author} {\bibinfo {author} {\bibfnamefont {S.}~\bibnamefont
  {Thirani}}, \bibinfo {author} {\bibfnamefont {P.}~\bibnamefont {Gupta}}, \
  and\ \bibinfo {author} {\bibfnamefont {C.}~\bibnamefont {Scalo}},\ }\bibfield
   {title} {\enquote {\bibinfo {title} {Knudsen number effects on the nonlinear
  acoustic spectral energy cascade},}\ }\href@noop {} {\bibfield  {journal}
  {\bibinfo  {journal} {Physical Review E}\ }\textbf {\bibinfo {volume}
  {101}},\ \bibinfo {pages} {023101} (\bibinfo {year} {2020})}\BibitemShut
  {NoStop}%
\bibitem [{\citenamefont {Ellermeier}(1993)}]{ellermeier1993nonlinear}%
  \BibitemOpen
  \bibfield  {author} {\bibinfo {author} {\bibfnamefont {W.}~\bibnamefont
  {Ellermeier}},\ }\bibfield  {title} {\enquote {\bibinfo {title} {Nonlinear
  acoustics in non-uniform infinite and finite layers},}\ }\href@noop {}
  {\bibfield  {journal} {\bibinfo  {journal} {Journal of Fluid Mechanics}\
  }\textbf {\bibinfo {volume} {257}},\ \bibinfo {pages} {183--200} (\bibinfo
  {year} {1993})}\BibitemShut {NoStop}%
\bibitem [{\citenamefont {Tyagi}\ and\ \citenamefont
  {Sujith}(2003)}]{tyagi2003nonlinear}%
  \BibitemOpen
  \bibfield  {author} {\bibinfo {author} {\bibfnamefont {M.}~\bibnamefont
  {Tyagi}}\ and\ \bibinfo {author} {\bibfnamefont {R.~I.}\ \bibnamefont
  {Sujith}},\ }\bibfield  {title} {\enquote {\bibinfo {title} {Nonlinear
  distortion of travelling waves in variable-area ducts with entropy
  gradients},}\ }\href@noop {} {\bibfield  {journal} {\bibinfo  {journal}
  {Journal of Fluid Mechanics}\ }\textbf {\bibinfo {volume} {492}},\ \bibinfo
  {pages} {1--22} (\bibinfo {year} {2003})}\BibitemShut {NoStop}%
\bibitem [{\citenamefont {Prasad}(2006)}]{prasad2006weakly}%
  \BibitemOpen
  \bibfield  {author} {\bibinfo {author} {\bibfnamefont {D.}~\bibnamefont
  {Prasad}},\ }\bibfield  {title} {\enquote {\bibinfo {title} {Weakly nonlinear
  shock propagation in slowly varying one-dimensional flows},}\ }\href@noop {}
  {\bibfield  {journal} {\bibinfo  {journal} {Physics of Fluids}\ }\textbf
  {\bibinfo {volume} {18}},\ \bibinfo {pages} {036101} (\bibinfo {year}
  {2006})}\BibitemShut {NoStop}%
\bibitem [{\citenamefont {Budzinsky}, \citenamefont {Zukoski},\ and\
  \citenamefont {Marble}(1992)}]{budzinski1992rayleigh}%
  \BibitemOpen
  \bibfield  {author} {\bibinfo {author} {\bibfnamefont {J.}~\bibnamefont
  {Budzinsky}}, \bibinfo {author} {\bibfnamefont {E.}~\bibnamefont {Zukoski}},
  \ and\ \bibinfo {author} {\bibfnamefont {F.}~\bibnamefont {Marble}},\
  }\bibfield  {title} {\enquote {\bibinfo {title} {Rayleigh scattering
  measurements of shock enhanced mixing},}\ }in\ \href@noop {} {\emph {\bibinfo
  {booktitle} {28th Joint Propulsion Conference and Exhibit}}}\ (\bibinfo
  {year} {1992})\ p.\ \bibinfo {pages} {3546}\BibitemShut {NoStop}%
\bibitem [{\citenamefont {Andreopoulos}, \citenamefont {Agui},\ and\
  \citenamefont {Briassulis}(2000)}]{andreopoulos2000shock}%
  \BibitemOpen
  \bibfield  {author} {\bibinfo {author} {\bibfnamefont {Y.}~\bibnamefont
  {Andreopoulos}}, \bibinfo {author} {\bibfnamefont {J.~H.}\ \bibnamefont
  {Agui}}, \ and\ \bibinfo {author} {\bibfnamefont {G.}~\bibnamefont
  {Briassulis}},\ }\bibfield  {title} {\enquote {\bibinfo {title} {Shock
  wave—turbulence interactions},}\ }\href@noop {} {\bibfield  {journal}
  {\bibinfo  {journal} {Annual Review of Fluid Mechanics}\ }\textbf {\bibinfo
  {volume} {32}},\ \bibinfo {pages} {309--345} (\bibinfo {year}
  {2000})}\BibitemShut {NoStop}%
\bibitem [{\citenamefont {Romagnosi}\ \emph {et~al.}(2011)\citenamefont
  {Romagnosi}, \citenamefont {Ingenito}, \citenamefont {Cecere}, \citenamefont
  {Eugenio},\ and\ \citenamefont {Bruno}}]{romagnosi2011role}%
  \BibitemOpen
  \bibfield  {author} {\bibinfo {author} {\bibfnamefont {L.}~\bibnamefont
  {Romagnosi}}, \bibinfo {author} {\bibfnamefont {A.}~\bibnamefont {Ingenito}},
  \bibinfo {author} {\bibfnamefont {D.}~\bibnamefont {Cecere}}, \bibinfo
  {author} {\bibfnamefont {G.}~\bibnamefont {Eugenio}}, \ and\ \bibinfo
  {author} {\bibfnamefont {C.}~\bibnamefont {Bruno}},\ }\bibfield  {title}
  {\enquote {\bibinfo {title} {The role of the baroclinic term in supersonic
  fuel/air mixing enhancement},}\ }in\ \href@noop {} {\emph {\bibinfo
  {booktitle} {49th AIAA Aerospace Sciences Meeting including the New Horizons
  Forum and Aerospace Exposition}}}\ (\bibinfo {year} {2011})\ p.\ \bibinfo
  {pages} {401}\BibitemShut {NoStop}%
\bibitem [{\citenamefont {Tian}\ \emph {et~al.}(2017)\citenamefont {Tian},
  \citenamefont {Jaberi}, \citenamefont {Li},\ and\ \citenamefont
  {Livescu}}]{tian2017numerical}%
  \BibitemOpen
  \bibfield  {author} {\bibinfo {author} {\bibfnamefont {Y.}~\bibnamefont
  {Tian}}, \bibinfo {author} {\bibfnamefont {F.~A.}\ \bibnamefont {Jaberi}},
  \bibinfo {author} {\bibfnamefont {Z.}~\bibnamefont {Li}}, \ and\ \bibinfo
  {author} {\bibfnamefont {D.}~\bibnamefont {Livescu}},\ }\bibfield  {title}
  {\enquote {\bibinfo {title} {Numerical study of variable density turbulence
  interaction with a normal shock wave},}\ }\href@noop {} {\bibfield  {journal}
  {\bibinfo  {journal} {Journal of Fluid Mechanics}\ }\textbf {\bibinfo
  {volume} {829}},\ \bibinfo {pages} {551--588} (\bibinfo {year}
  {2017})}\BibitemShut {NoStop}%
\bibitem [{\citenamefont {Wong}\ \emph {et~al.}(2022)\citenamefont {Wong},
  \citenamefont {Baltzer}, \citenamefont {Livescu},\ and\ \citenamefont
  {Lele}}]{wong2022analysis}%
  \BibitemOpen
  \bibfield  {author} {\bibinfo {author} {\bibfnamefont {M.~L.}\ \bibnamefont
  {Wong}}, \bibinfo {author} {\bibfnamefont {J.~R.}\ \bibnamefont {Baltzer}},
  \bibinfo {author} {\bibfnamefont {D.}~\bibnamefont {Livescu}}, \ and\
  \bibinfo {author} {\bibfnamefont {S.~K.}\ \bibnamefont {Lele}},\ }\bibfield
  {title} {\enquote {\bibinfo {title} {Analysis of second moments and their
  budgets for richtmyer-meshkov instability and variable-density turbulence
  induced by reshock},}\ }\href@noop {} {\bibfield  {journal} {\bibinfo
  {journal} {Physical Review Fluids}\ }\textbf {\bibinfo {volume} {7}},\
  \bibinfo {pages} {044602} (\bibinfo {year} {2022})}\BibitemShut {NoStop}%
\bibitem [{\citenamefont {Singh}\ \emph {et~al.}(2019)\citenamefont {Singh},
  \citenamefont {Rajendran}, \citenamefont {Gupta}, \citenamefont {Scalo},
  \citenamefont {Vlachos},\ and\ \citenamefont {Bane}}]{singh2019experimental}%
  \BibitemOpen
  \bibfield  {author} {\bibinfo {author} {\bibfnamefont {B.}~\bibnamefont
  {Singh}}, \bibinfo {author} {\bibfnamefont {L.~K.}\ \bibnamefont
  {Rajendran}}, \bibinfo {author} {\bibfnamefont {P.}~\bibnamefont {Gupta}},
  \bibinfo {author} {\bibfnamefont {C.}~\bibnamefont {Scalo}}, \bibinfo
  {author} {\bibfnamefont {P.~P.}\ \bibnamefont {Vlachos}}, \ and\ \bibinfo
  {author} {\bibfnamefont {S.~P.}\ \bibnamefont {Bane}},\ }\bibfield  {title}
  {\enquote {\bibinfo {title} {Experimental and numerical study of flow induced
  by nanosecond repetitively pulsed discharges},}\ }in\ \href@noop {} {\emph
  {\bibinfo {booktitle} {AIAA Scitech 2019 Forum}}}\ (\bibinfo {year} {2019})\
  p.\ \bibinfo {pages} {0740}\BibitemShut {NoStop}%
\bibitem [{\citenamefont {Kundu}, \citenamefont {Cohen},\ and\ \citenamefont
  {Dowling}(2015)}]{kundu2015fluid}%
  \BibitemOpen
  \bibfield  {author} {\bibinfo {author} {\bibfnamefont {P.~K.}\ \bibnamefont
  {Kundu}}, \bibinfo {author} {\bibfnamefont {I.~M.}\ \bibnamefont {Cohen}}, \
  and\ \bibinfo {author} {\bibfnamefont {D.~R.}\ \bibnamefont {Dowling}},\
  }\href@noop {} {\emph {\bibinfo {title} {Fluid mechanics}}}\ (\bibinfo
  {publisher} {Academic press},\ \bibinfo {year} {2015})\BibitemShut {NoStop}%
\bibitem [{\citenamefont {Mikaelian}(1985)}]{mikaelian1985richtmyer}%
  \BibitemOpen
  \bibfield  {author} {\bibinfo {author} {\bibfnamefont {K.~O.}\ \bibnamefont
  {Mikaelian}},\ }\bibfield  {title} {\enquote {\bibinfo {title}
  {Richtmyer-meshkov instabilities in stratified fluids},}\ }\href@noop {}
  {\bibfield  {journal} {\bibinfo  {journal} {Physical Review A}\ }\textbf
  {\bibinfo {volume} {31}},\ \bibinfo {pages} {410} (\bibinfo {year}
  {1985})}\BibitemShut {NoStop}%
\bibitem [{\citenamefont {Yang}\ and\ \citenamefont
  {Radulescu}(2021)}]{yang2021dynamics}%
  \BibitemOpen
  \bibfield  {author} {\bibinfo {author} {\bibfnamefont {H.}~\bibnamefont
  {Yang}}\ and\ \bibinfo {author} {\bibfnamefont {M.~I.}\ \bibnamefont
  {Radulescu}},\ }\bibfield  {title} {\enquote {\bibinfo {title} {Dynamics of
  cellular flame deformation after a head-on interaction with a shock wave:
  reactive richtmyer--meshkov instability},}\ }\href@noop {} {\bibfield
  {journal} {\bibinfo  {journal} {Journal of Fluid Mechanics}\ }\textbf
  {\bibinfo {volume} {923}},\ \bibinfo {pages} {A36} (\bibinfo {year}
  {2021})}\BibitemShut {NoStop}%
\bibitem [{\citenamefont {Yu}\ \emph {et~al.}(2020)\citenamefont {Yu},
  \citenamefont {He}, \citenamefont {Zhang},\ and\ \citenamefont
  {Liu}}]{yu2020two}%
  \BibitemOpen
  \bibfield  {author} {\bibinfo {author} {\bibfnamefont {B.}~\bibnamefont
  {Yu}}, \bibinfo {author} {\bibfnamefont {M.}~\bibnamefont {He}}, \bibinfo
  {author} {\bibfnamefont {B.}~\bibnamefont {Zhang}}, \ and\ \bibinfo {author}
  {\bibfnamefont {H.}~\bibnamefont {Liu}},\ }\bibfield  {title} {\enquote
  {\bibinfo {title} {Two-stage growth mode for lift-off mechanism in oblique
  shock-wave/jet interaction},}\ }\href@noop {} {\bibfield  {journal} {\bibinfo
   {journal} {Physics of Fluids}\ }\textbf {\bibinfo {volume} {32}},\ \bibinfo
  {pages} {116105} (\bibinfo {year} {2020})}\BibitemShut {NoStop}%
\bibitem [{\citenamefont {Wei}\ \emph {et~al.}(2022)\citenamefont {Wei},
  \citenamefont {Yang}, \citenamefont {Liu}, \citenamefont {Zhao},
  \citenamefont {Wang},\ and\ \citenamefont {Sun}}]{wei2022flow}%
  \BibitemOpen
  \bibfield  {author} {\bibinfo {author} {\bibfnamefont {F.}~\bibnamefont
  {Wei}}, \bibinfo {author} {\bibfnamefont {R.}~\bibnamefont {Yang}}, \bibinfo
  {author} {\bibfnamefont {W.}~\bibnamefont {Liu}}, \bibinfo {author}
  {\bibfnamefont {Y.}~\bibnamefont {Zhao}}, \bibinfo {author} {\bibfnamefont
  {Q.}~\bibnamefont {Wang}}, \ and\ \bibinfo {author} {\bibfnamefont
  {M.}~\bibnamefont {Sun}},\ }\bibfield  {title} {\enquote {\bibinfo {title}
  {Flow structures of strong interaction between an oblique shock wave and a
  supersonic streamwise vortex},}\ }\href@noop {} {\bibfield  {journal}
  {\bibinfo  {journal} {Physics of Fluids}\ }\textbf {\bibinfo {volume} {34}},\
  \bibinfo {pages} {106102} (\bibinfo {year} {2022})}\BibitemShut {NoStop}%
\bibitem [{\citenamefont {Eswaran}\ and\ \citenamefont
  {Pope}(1988)}]{eswaran1988examination}%
  \BibitemOpen
  \bibfield  {author} {\bibinfo {author} {\bibfnamefont {V.}~\bibnamefont
  {Eswaran}}\ and\ \bibinfo {author} {\bibfnamefont {S.~B.}\ \bibnamefont
  {Pope}},\ }\bibfield  {title} {\enquote {\bibinfo {title} {An examination of
  forcing in direct numerical simulations of turbulence},}\ }\href@noop {}
  {\bibfield  {journal} {\bibinfo  {journal} {Computers \& Fluids}\ }\textbf
  {\bibinfo {volume} {16}},\ \bibinfo {pages} {257--278} (\bibinfo {year}
  {1988})}\BibitemShut {NoStop}%
\bibitem [{\citenamefont {Jagannathan}\ and\ \citenamefont
  {Donzis}(2016)}]{jagannathan2016reynolds}%
  \BibitemOpen
  \bibfield  {author} {\bibinfo {author} {\bibfnamefont {S.}~\bibnamefont
  {Jagannathan}}\ and\ \bibinfo {author} {\bibfnamefont {D.~A.}\ \bibnamefont
  {Donzis}},\ }\bibfield  {title} {\enquote {\bibinfo {title} {Reynolds and
  mach number scaling in solenoidally-forced compressible turbulence using
  high-resolution direct numerical simulations},}\ }\href@noop {} {\bibfield
  {journal} {\bibinfo  {journal} {Journal of Fluid Mechanics}\ }\textbf
  {\bibinfo {volume} {789}},\ \bibinfo {pages} {669--707} (\bibinfo {year}
  {2016})}\BibitemShut {NoStop}%
\bibitem [{\citenamefont {Miura}\ and\ \citenamefont
  {Kida}(1995)}]{miura1995acoustic}%
  \BibitemOpen
  \bibfield  {author} {\bibinfo {author} {\bibfnamefont {H.}~\bibnamefont
  {Miura}}\ and\ \bibinfo {author} {\bibfnamefont {S.}~\bibnamefont {Kida}},\
  }\bibfield  {title} {\enquote {\bibinfo {title} {Acoustic energy exchange in
  compressible turbulence},}\ }\href@noop {} {\bibfield  {journal} {\bibinfo
  {journal} {Physics of Fluids}\ }\textbf {\bibinfo {volume} {7}},\ \bibinfo
  {pages} {1732--1742} (\bibinfo {year} {1995})}\BibitemShut {NoStop}%
\bibitem [{\citenamefont {Boyd}(2001)}]{boyd2001chebyshev}%
  \BibitemOpen
  \bibfield  {author} {\bibinfo {author} {\bibfnamefont {J.~P.}\ \bibnamefont
  {Boyd}},\ }\href@noop {} {\emph {\bibinfo {title} {Chebyshev and Fourier
  spectral methods}}}\ (\bibinfo  {publisher} {Courier Corporation},\ \bibinfo
  {year} {2001})\BibitemShut {NoStop}%
\bibitem [{\citenamefont {Mortensen}\ and\ \citenamefont
  {Langtangen}(2016)}]{mortensen2016high}%
  \BibitemOpen
  \bibfield  {author} {\bibinfo {author} {\bibfnamefont {M.}~\bibnamefont
  {Mortensen}}\ and\ \bibinfo {author} {\bibfnamefont {H.~P.}\ \bibnamefont
  {Langtangen}},\ }\bibfield  {title} {\enquote {\bibinfo {title} {High
  performance python for direct numerical simulations of turbulent flows},}\
  }\href@noop {} {\bibfield  {journal} {\bibinfo  {journal} {Computer Physics
  Communications}\ }\textbf {\bibinfo {volume} {203}},\ \bibinfo {pages}
  {53--65} (\bibinfo {year} {2016})}\BibitemShut {NoStop}%
\bibitem [{\citenamefont {Yeung}, \citenamefont {Sreenivasan},\ and\
  \citenamefont {Pope}(2018)}]{yeung2018effects}%
  \BibitemOpen
  \bibfield  {author} {\bibinfo {author} {\bibfnamefont {P.}~\bibnamefont
  {Yeung}}, \bibinfo {author} {\bibfnamefont {K.}~\bibnamefont {Sreenivasan}},
  \ and\ \bibinfo {author} {\bibfnamefont {S.}~\bibnamefont {Pope}},\
  }\bibfield  {title} {\enquote {\bibinfo {title} {Effects of finite spatial
  and temporal resolution in direct numerical simulations of incompressible
  isotropic turbulence},}\ }\href@noop {} {\bibfield  {journal} {\bibinfo
  {journal} {Physical Review Fluids}\ }\textbf {\bibinfo {volume} {3}},\
  \bibinfo {pages} {064603} (\bibinfo {year} {2018})}\BibitemShut {NoStop}%
\bibitem [{\citenamefont {Bell}\ \emph {et~al.}(2022)\citenamefont {Bell},
  \citenamefont {Nonaka}, \citenamefont {Garcia},\ and\ \citenamefont
  {Eyink}}]{bell2022thermal}%
  \BibitemOpen
  \bibfield  {author} {\bibinfo {author} {\bibfnamefont {J.~B.}\ \bibnamefont
  {Bell}}, \bibinfo {author} {\bibfnamefont {A.}~\bibnamefont {Nonaka}},
  \bibinfo {author} {\bibfnamefont {A.~L.}\ \bibnamefont {Garcia}}, \ and\
  \bibinfo {author} {\bibfnamefont {G.}~\bibnamefont {Eyink}},\ }\bibfield
  {title} {\enquote {\bibinfo {title} {Thermal fluctuations in the dissipation
  range of homogeneous isotropic turbulence},}\ }\href@noop {} {\bibfield
  {journal} {\bibinfo  {journal} {Journal of Fluid Mechanics}\ }\textbf
  {\bibinfo {volume} {939}},\ \bibinfo {pages} {A12} (\bibinfo {year}
  {2022})}\BibitemShut {NoStop}%
\bibitem [{\citenamefont {J~Evans}\ and\ \citenamefont
  {P~Morriss}(2007)}]{j2007statistical}%
  \BibitemOpen
  \bibfield  {author} {\bibinfo {author} {\bibfnamefont {D.}~\bibnamefont
  {J~Evans}}\ and\ \bibinfo {author} {\bibfnamefont {G.}~\bibnamefont
  {P~Morriss}},\ }\href@noop {} {\emph {\bibinfo {title} {Statistical mechanics
  of nonequilbrium liquids}}}\ (\bibinfo  {publisher} {ANU Press},\ \bibinfo
  {year} {2007})\BibitemShut {NoStop}%
\bibitem [{\citenamefont {Ghosal}\ \emph {et~al.}(1995)\citenamefont {Ghosal},
  \citenamefont {Lund}, \citenamefont {Moin},\ and\ \citenamefont
  {Akselvoll}}]{ghosal1995dynamic}%
  \BibitemOpen
  \bibfield  {author} {\bibinfo {author} {\bibfnamefont {S.}~\bibnamefont
  {Ghosal}}, \bibinfo {author} {\bibfnamefont {T.~S.}\ \bibnamefont {Lund}},
  \bibinfo {author} {\bibfnamefont {P.}~\bibnamefont {Moin}}, \ and\ \bibinfo
  {author} {\bibfnamefont {K.}~\bibnamefont {Akselvoll}},\ }\bibfield  {title}
  {\enquote {\bibinfo {title} {A dynamic localization model for large-eddy
  simulation of turbulent flows},}\ }\href@noop {} {\bibfield  {journal}
  {\bibinfo  {journal} {Journal of fluid mechanics}\ }\textbf {\bibinfo
  {volume} {286}},\ \bibinfo {pages} {229--255} (\bibinfo {year}
  {1995})}\BibitemShut {NoStop}%
\bibitem [{\citenamefont {Augier}, \citenamefont {Mohanan},\ and\ \citenamefont
  {Lindborg}(2019)}]{augier2019shallow}%
  \BibitemOpen
  \bibfield  {author} {\bibinfo {author} {\bibfnamefont {P.}~\bibnamefont
  {Augier}}, \bibinfo {author} {\bibfnamefont {A.~V.}\ \bibnamefont {Mohanan}},
  \ and\ \bibinfo {author} {\bibfnamefont {E.}~\bibnamefont {Lindborg}},\
  }\bibfield  {title} {\enquote {\bibinfo {title} {Shallow water wave
  turbulence},}\ }\href@noop {} {\bibfield  {journal} {\bibinfo  {journal}
  {Journal of Fluid Mechanics}\ }\textbf {\bibinfo {volume} {874}},\ \bibinfo
  {pages} {1169--1196} (\bibinfo {year} {2019})}\BibitemShut {NoStop}%
\bibitem [{\citenamefont {Verma}(2019)}]{verma2019energy}%
  \BibitemOpen
  \bibfield  {author} {\bibinfo {author} {\bibfnamefont {M.~K.}\ \bibnamefont
  {Verma}},\ }\href@noop {} {\emph {\bibinfo {title} {Energy transfers in fluid
  flows: multiscale and spectral perspectives}}}\ (\bibinfo  {publisher}
  {Cambridge University Press},\ \bibinfo {year} {2019})\BibitemShut {NoStop}%
\bibitem [{\citenamefont {Verma}(2021)}]{verma2021variable}%
  \BibitemOpen
  \bibfield  {author} {\bibinfo {author} {\bibfnamefont {M.~K.}\ \bibnamefont
  {Verma}},\ }\bibfield  {title} {\enquote {\bibinfo {title} {Variable energy
  flux in turbulence},}\ }\href@noop {} {\bibfield  {journal} {\bibinfo
  {journal} {Journal of Physics A: Mathematical and Theoretical}\ }\textbf
  {\bibinfo {volume} {55}},\ \bibinfo {pages} {013002} (\bibinfo {year}
  {2021})}\BibitemShut {NoStop}%
\bibitem [{\citenamefont {Pope}(2000)}]{pope2000turbulent}%
  \BibitemOpen
  \bibfield  {author} {\bibinfo {author} {\bibfnamefont {S.~B.}\ \bibnamefont
  {Pope}},\ }\href@noop {} {\emph {\bibinfo {title} {Turbulent flows}}}\
  (\bibinfo  {publisher} {Cambridge university press},\ \bibinfo {year}
  {2000})\BibitemShut {NoStop}%
\bibitem [{\citenamefont {Kida}\ and\ \citenamefont
  {Orszag}(1990)}]{kida1990enstrophy}%
  \BibitemOpen
  \bibfield  {author} {\bibinfo {author} {\bibfnamefont {S.}~\bibnamefont
  {Kida}}\ and\ \bibinfo {author} {\bibfnamefont {S.~A.}\ \bibnamefont
  {Orszag}},\ }\bibfield  {title} {\enquote {\bibinfo {title} {Enstrophy budget
  in decaying compressible turbulence},}\ }\href@noop {} {\bibfield  {journal}
  {\bibinfo  {journal} {Journal of scientific computing}\ }\textbf {\bibinfo
  {volume} {5}},\ \bibinfo {pages} {1--34} (\bibinfo {year}
  {1990})}\BibitemShut {NoStop}%
\bibitem [{\citenamefont {Dowling}\ and\ \citenamefont
  {Mahmoudi}(2015)}]{dowling2015combustion}%
  \BibitemOpen
  \bibfield  {author} {\bibinfo {author} {\bibfnamefont {A.~P.}\ \bibnamefont
  {Dowling}}\ and\ \bibinfo {author} {\bibfnamefont {Y.}~\bibnamefont
  {Mahmoudi}},\ }\bibfield  {title} {\enquote {\bibinfo {title} {Combustion
  noise},}\ }\href@noop {} {\bibfield  {journal} {\bibinfo  {journal}
  {Proceedings of the Combustion Institute}\ }\textbf {\bibinfo {volume}
  {35}},\ \bibinfo {pages} {65--100} (\bibinfo {year} {2015})}\BibitemShut
  {NoStop}%
\bibitem [{\citenamefont {Nazarenko}(2011)}]{nazarenko2011wave}%
  \BibitemOpen
  \bibfield  {author} {\bibinfo {author} {\bibfnamefont {S.}~\bibnamefont
  {Nazarenko}},\ }\href@noop {} {\emph {\bibinfo {title} {Wave turbulence}}},\
  Vol.\ \bibinfo {volume} {825}\ (\bibinfo  {publisher} {Springer Science \&
  Business Media},\ \bibinfo {year} {2011})\BibitemShut {NoStop}%
\bibitem [{\citenamefont {Rosales}\ and\ \citenamefont
  {Meneveau}(2005)}]{rosales2005linear}%
  \BibitemOpen
  \bibfield  {author} {\bibinfo {author} {\bibfnamefont {C.}~\bibnamefont
  {Rosales}}\ and\ \bibinfo {author} {\bibfnamefont {C.}~\bibnamefont
  {Meneveau}},\ }\bibfield  {title} {\enquote {\bibinfo {title} {Linear forcing
  in numerical simulations of isotropic turbulence: Physical space
  implementations and convergence properties},}\ }\href@noop {} {\bibfield
  {journal} {\bibinfo  {journal} {Physics of fluids}\ }\textbf {\bibinfo
  {volume} {17}},\ \bibinfo {pages} {095106} (\bibinfo {year}
  {2005})}\BibitemShut {NoStop}%
\bibitem [{\citenamefont {Farge}\ and\ \citenamefont
  {Sadourny}(1989)}]{farge1989wave}%
  \BibitemOpen
  \bibfield  {author} {\bibinfo {author} {\bibfnamefont {M.}~\bibnamefont
  {Farge}}\ and\ \bibinfo {author} {\bibfnamefont {R.}~\bibnamefont
  {Sadourny}},\ }\bibfield  {title} {\enquote {\bibinfo {title} {Wave-vortex
  dynamics in rotating shallow water},}\ }\href@noop {} {\bibfield  {journal}
  {\bibinfo  {journal} {Journal of Fluid Mechanics}\ }\textbf {\bibinfo
  {volume} {206}},\ \bibinfo {pages} {433--462} (\bibinfo {year}
  {1989})}\BibitemShut {NoStop}%
\end{thebibliography}%
